\documentclass[aps,pre,preprint,groupedaddress,showpacs]{revtex4}

\begin{document}

\title{Interaction of a soliton with a localized gain in a fiber Bragg grating}
\author{William C. K. Mak$^{1}$, Boris A. Malomed$^{2,1}$, and Pak L. Chu$^1$}

\affiliation{$^1$Optoelectronic Research Centre, Department of Electronic
Engineering, City University of Hong Kong, Tat Chee Avenue, Kowloon, Hong Kong}

\affiliation{$^2$Department of Interdisciplinary Studies,
Faculty of Engineering, Tel Aviv University,
Tel Aviv 69978, Israel}

\begin{abstract}
A model of a lossy nonlinear fiber grating with a ``hot spot'', which
combines a local gain and an attractive perturbation of the refractive
index, is introduced. A family of exact solutions for pinned solitons is
found in the absence of loss and gain. In the presence of the loss and
localized gain, an instability threshold of the zero solution is found. If
the loss and gain are small, it is predicted what soliton is selected by the
energy-balance condition. Direct simulations demonstrate that only one
pinned soliton is stable in the conservative model, and it is a $\mathit{\
semi-attractor}$: solitons with a larger energy relax to it via emission of
radiation, while those with a smaller energy decay. The same is found for
solitons trapped by a pair of repulsive inhomogeneities. In the model with
the loss and gain, stable pinned pulses demonstrate persistent internal
vibrations and emission of radiation. If these solitons are nearly
stationary, the prediction based on the energy balance underestimates the
necessary gain by $10-15\%$ (due to radiation loss). If the loss and gain
are larger, the intrinsic vibrations of the pinned soliton become chaotic.
The local gain alone, without the attractive perturbation of the local
refractive index, cannot maintain a stable pinned soliton. For collisions of
moving solitons with the ``hot spot'', passage and capture regimes are
identified, the capture actually implying splitting of the soliton.
\end{abstract}

\maketitle

\section{Introduction}

Solitons in any physical medium are subject to attenuation due to
dissipation, hence it is necessary to apply gain which can support the
solitons. A possibility which may find interesting physical applications is
to create a localized gain, which will be a trap for solitons in a lossy
medium. In terms of optical solitons, this can be easily realized for
spatial solitons in a planar waveguide, where the gain may be applied to a
narrow strip. However, for spatial solitons gain is a redundancy, as they
are supported simply by the energy flux through the soliton itself. Besides
that, in this work it will be shown that, in the simplest case when the
spatial solitons are governed by the nonlinear Schr\"{o}dinger (NLS)
equation, a pinned soliton supported by the localized gain can never be
stable. On the other hand, realization of the local-gain trap for usual
temporal solitons is impossible, as such a soliton runs (for instance, in an
optical fiber \cite{Agrawal}) with the group velocity of light.

An unique possibility to create a gain-induced trap in a lossy medium is
offered by a fiber grating, i.e., a Bragg grating (BG) written on an optical
fiber. Fiber gratings are a basis for many photonic devices \cite{Kashyap}.
A challenging possibility is to use fiber gratings for the creation of
pulses of slow light, which is a topic of great current interest 
\cite{slow}. The possibility of the existence of slow pulses 
suggest to try a local-gain-induced trap for solitons.

In fiber gratings, solitons exist due to the interplay between the Bragg
reflection and Kerr nonlinearity of the BG-carrying fiber \cite{Sterke}.
These solitons were predicted analytically \cite{Aceves,Demetri}, and then
they were created in the experiment \cite{experiment}. Except for the case
in which the BG soliton is very broad \cite{Litchinitser}, this species of
optical solitons is distinct from the usual nonlinear-Schr\"{o}dinger (NLS)
solitons \cite{Agrawal} in ordinary nonlinear optical fibers.

As it was mentioned above, search for very slow solitons in fiber gratings,
which were predicted long ago \cite{Demetri}, is an issue of great interest 
\cite{slowsoliton}. The standard mathematical model of the nonlinear fiber
grating predicts a whole family of zero-velocity solitons \cite
{Aceves,Demetri,Sterke}, a part of which is stable \cite{Rich,Barash}. If
realized experimentally, such a soliton would represent a pulse of standing
light, with its left- and right-traveling components being in a permanent
dynamical equilibrium.

The BG solitons which have been observed in the experiment up to date are
moving ones, their velocity being $\simeq 75\%$ of the maximum group
velocity of light in the fiber \cite{experiment}. The quest for
zero-velocity solitons may be facilitated by means of a local defect in the
BG that exerts an attractive force on a soliton, having thus a potential to
be a soliton trap (note that a local defect may trap light in the fiber
grating via the four-wave mixing without formation of a soliton \cite
{trapping}). Besides its profound physical importance \cite{Weinstein}, such
a soliton trap is also promising for the fiber-sensing technology \cite{kdv}.

The interaction of the soliton with an attractive defect in the form of a
local suppression of BG was studied recently in Refs. \cite{Weinstein} and 
\cite{kdv}. The latter work also considered a trap combining the local BG
suppression and a change in the refractive index (these two effects may come
together as a manifestation of a local inhomogeneity in the BG-carrying
fiber). As a result, it has been demonstrated that a gap soliton may indeed
be pinned by the local defect, and, moreover, a narrow (delta-functional)
defect uniquely selects parameters of the stable trapped soliton \cite{kdv}
(this feature is described in more detail below).

However, a trapped soliton will be destroyed by fiber loss. Indeed, taking
into regard that the best fiber gratings (used as dispersion compensators in
optical telecommunications \cite{Krug}) have the attenuation rate $\simeq
0.2 $ dB/cm (in hybrid grating waveguides, using glass and a sol-gel
material, the attenuation may be lowered to $0.1$ dB/cm \cite{solgel}), it
is easy to estimate that a standing soliton will be destroyed during the
time $_{\sim }^{<}\,5$ ns. Therefore, it is necessary to support the trapped
soliton by means of a locally applied optical gain, which is tantamount to
the above-mentioned soliton trap induced by local gain. A related problem is
a possibility to capture a moving soliton by the local gain. These issues
are the main subjects of the present work. Besides the fundamental interest
to having permanently maintained trapped optical solitons, they may also be
interesting for applications, such as optical memory.

The local gain can be provided, for instance, by a short resonantly doped
segment in the BG-carrying fiber. In this connection, we note that the
moving BG soliton that was observed for the first time in the fiber grating
had the temporal width $\simeq 200$ ps \cite{experiment}. Taking into regard
the Lorentzian contraction (as it was mentioned above, the soliton was
moving at a velocity equal $\approx 75\%$ of the maximum group velocity),
the same soliton, if stopped, would have the spatial width on the order of a
few centimeters. Thus, the locally pumped segment of the fiber may be
approximated by a delta-function (which we assume in the analytical model
below) if, roughly speaking, its size is $\simeq 1$ mm. Note that the
maximum gain that can be provided by an \textrm{Er}-doped amplifier is 
$\simeq 5$ dB/cm \cite{Erbium}. Comparing this to the above-mentioned minimum
damping rate $0.1$ dB/cm, we conclude that the proposed scheme may be
self-consistent.

Thus, the model to be formulated below assumes the local gain in the form of
the delta-function (in the numerical part of the work, a smooth
approximation to the delta-function is used). The model also includes an
imaginary part of the localized gain, which actually accounts for a local
perturbation of the fiber's refractive index (note that the dopant, if used
to induce the local gain, may indeed affect the local refractive index). The
imaginary part is necessary, as it will be seen that pure gain cannot
maintain a pinned soliton in a stable state. As for the size of the
refractive-index perturbation $\delta n$, it will be seen that $\delta
n\,\,_{\sim }^{>}\,\,0.1$ is definitely sufficient to stably trap a soliton.
In this respect, it is relevant to mention that a dopant added to a silica
fiber usually induces a refractive index change $\delta n\sim 0.05$, while
in a polymer fiber it is $\sim 0.2$. The latter value can readily produce a
stable pinned state of the soliton, and the former one may be sufficient too.

A relatively large jump of the refractive index between the doped segment
and the rest of the fiber may induce an additional effect, viz., reflection
of light, although the reflection may be smothered by a sufficiently smooth
profile of distribution of the dopant. An additional term in the basic model
[see Eqs. (\ref{o1}) and (\ref{o2}) below], induced by the reflection, would
be the same which formally accounts for a local perturbation of the BG
reflectivity. The latter type of the conservative perturbation was
considered in Ref. {\cite{kdv}), a conclusion being that its effect is quite
similar to that directly produced by the refractive-index perturbation,
which is directly included in Eqs. (\ref{o1}) and (\ref{o2}). For this
reason, we do not consider the local reflection as a separate perturbation
in this work (in any case, it can be easily added, if necessary). }

The paper is organized as follows. In section 2 we formulate the model, and
produce some analytical results. First of all, we consider the stability of
the zero solution against small perturbations in the presence of the uniform
loss and localized loss. An instability-onset threshold is found, and it is
demonstrated that the instability does not take place unless the gain has an
imaginary part. For the comparison's sake, we also briefly consider an
allied model problem, viz., the instability induced by a ``hot spot'' in the
lossy NLS equation, which produces quite similar results. Then, we produce a
family of exact analytical solutions for solitons pinned by the local
inhomogeneity of the refractive index, in the absence of loss and gain.
Another analytical result predicts what particular pinned soliton is
selected by the balance of the loss and gain, provided that they are small
perturbations. A soliton selected by the energy balance is also found in the
lossy NLS equation with the hot spot (in that case, it is always unstable).

Section 3 reports results of direct simulations of pinned solitons, in both
the conservative version of the model, and in the full one. A surprising
result is that, in the conservative model, a \emph{single} pinned soliton,
out of their continuous family, is stable. If the initially created soliton
has larger energy, it relaxes, via emission of radiation, to the stable one,
and if the energy of the initial soliton is smaller than that of the stable
one, the pulse decays into radiation; the former observation can be
explained by known results for the stability of the ordinary solitons in the
unperturbed BG model \cite{Rich,Barash}. A similar result is also obtained,
in the framework of the conservative model, at the end of section 3 for a
pair of repulsive defects: in the case when they can hold a soliton at the
midpoint between them, any pulse either relaxes to a uniquely selected
soliton, or decays (if the pair of repulsive inhomogeneities cannot hold a
soliton, it splits the soliton in two). It is relevant to mention that a
nonlinear BG structure with two local defects was very recently studied as a
promising model for all-optical switching \cite{Tran}.

In the full model, all the pinned solitons demonstrate persistent intrinsic
vibrations; depending on the ratio of the loss and gain, and on the strength
of the attractive refractive-index inhomogeneity, the amplitude of the
vibrations may be small or large. In the former case, the pinned soliton may
be regarded as an approximately stationary one, then the above-mentioned
analytical prediction of the soliton selected by the balance between the
loss and gain gives an error between $10\%$ and $15\%$, which may be
explained by extra radiation loss.

In section 4, collision of a moving soliton with the hot spot is considered
by means of direct simulations. As a result, regions of passage and capture
are identified in the soliton's parameter space in both the conservative and
full models. In fact, the capture is incomplete: a part of the soliton's
energy gets trapped, giving rise to a pinned soliton, while the other part
passes and rearranges itself into another soliton. If the inhomogeneity is
strong, a conspicuous part of the energy may bounce back.

\section{The model and analytical results}

\subsection{The model equations}

A localized gain inserted into a fiber grating is modelled by a 
$\delta$~-~function term added to the standard BG model, 
which includes coupled
equations for the amplitudes of the right- and left-travelling
electromagnetic waves, $u(x,t)$ and $v(x,t)$ \cite{Sterke}. The full model
is then

\begin{eqnarray}
iu_{t}+iu_{x}+v+\left( |u|^{2}+2|v|^{2}\right) u &=&-i\gamma u+i\Gamma
\delta (x)\cdot u,  \label{o1} \\
iv_{t}-iv_{x}+u+\left( |v|^{2}+2|u|^{2}\right) v &=&-i\gamma v+i\Gamma
\delta (x)\cdot v,  \label{o2}
\end{eqnarray}
where the maximum group velocity of light is normalized to be $1$, the
nonlinear terms account for the self- and cross-phase modulation induced by
the Kerr effect, the linear couplings represent the mutual conversion of the
waves due to the resonant Bragg scattering (the conversion coefficient is
also normalized to be $1$). On the right-hand side of Eqs. (\ref{o1}) and 
(\ref{o2}), the fiber loss parameter $\gamma $ is real and positive, while
the local-gain strength $\Gamma $ may be complex, 
\begin{equation}
\Gamma \equiv \Gamma _{1}+i\Gamma _{2}\,,  \label{Gamma}
\end{equation}
its positive real part being the gain proper, while the imaginary part
accounts for a localized perturbation of the refractive index (note that 
$\Gamma _{2}>0$ corresponds to a local increase of the refractive index).

A stationary solution, which represents a soliton trapped by the ``hot
spot'', is sought for as 
\begin{equation}
u(x,t)=U(x)\exp \left( -it\,\cos \theta \right) ,\,v(x,t)=V(x)\exp \left(
-it\,\cos \theta \right) ,  \label{statasmpt}
\end{equation}
where $\theta $ is a parameter of the soliton family. The substitution of
Eqs. (\ref{statasmpt}) into Eqs. (\ref{o1}) and (\ref{o2}) leads to
equations 
\begin{eqnarray}
\left[ i\frac{d}{dx}+\cos \theta +i\gamma +i\Gamma \delta (x)+\left(
|U|^{2}+2|V|^{2}\right) \right] U+V &=&0,  \label{U} \\
\left[ -i\frac{d}{dx}+\cos \theta +i\gamma +i\Gamma \delta (x)+\left(
|V|^{2}+2|U|^{2}\right) \right] V+U &=&0.  \label{V}
\end{eqnarray}

\subsection{Stability of the zero solution}

We start the analysis with the linearized version of Eqs.~(\ref{o1}) and 
(\ref{o2}), in order to analyze the stability of the zero solution in the
presence of the localized gain. An eigenmode of small perturbations is sough
for as

\begin{eqnarray}
u(x,t)=A_{+}\,\mathrm{exp}(-i\chi t-\kappa x) &,&\,v(x,t)=B_{+}\,\mathrm{exp}
(-i\chi t-\kappa x)\;\;\;\;\mathrm{at}\;\;\;\;x>0,  \label{lin1} \\
u(x,t)=A_{-}\,\mathrm{exp}(-i\chi t+\kappa x) &,&\,v(x,t)=B_{-}\,\mathrm{exp}
(-i\chi t+\kappa x)\;\;\;\;\mathrm{at}\;\;\;\;x<0,  \label{lin2}
\end{eqnarray}
where the frequency $\chi $ may be complex, its imaginary part being the
instability growth rate, and $\kappa $ must have a positive real part.
Substituting this into the linearized equations yields 
\begin{equation}
\kappa =\sqrt{1-(\chi +i\gamma )^{2}},  \label{kog}
\end{equation}
\noindent where the square root is defined so that its real part is
positive, and $B_{\pm }=-\left[ (\chi +i\gamma )\mp i\kappa \right] A_{\pm }$.
Further, the integration of the linearized equations in an infinitesimal
vicinity of $x=0$ yields $A_{+}=A_{-}\,e^{\Gamma },\;B_{+}=B_{-}\,e^{-\Gamma
}$.

Eliminating the amplitudes $A_{\pm }$ and $B_{\pm }$ by means of these
relations, we obtain an equation 
\begin{equation}
\frac{\chi +i\gamma -i\kappa }{\chi +i\gamma +i\kappa }=e^{-2\left( \Gamma
_{1}+i\Gamma _{2}\right) },  \label{kappa}
\end{equation}
where the expression (\ref{Gamma}) was substituted for $\Gamma $. Finally,
combining Eqs. (\ref{kappa}) and (\ref{kog}), we can find the eigenfrequency
sought for: 
\begin{equation}
\chi =-i\gamma +\mathrm{sgn}\left( \sin \Gamma _{2}\right) \cdot \cosh
\left( \Gamma _{1}+i\Gamma _{2}\right) \,,  \label{chi}
\end{equation}
where the sign multiplier in front of the second term provides for the
fulfillment of the above-mentioned condition $\mathrm{Re}\,\kappa >0$.

A straightforward consequence of Eq.~(\ref{chi}) is an expression for the
instability growth rate, 
\begin{equation}
\mathrm{Im}\,\chi =-\gamma +\left( \sinh \,\Gamma _{1}\right) \,|\sin
\,\Gamma _{2}|\,.  \label{linrel3}
\end{equation}
Thus, the localized gain gives rise to the instability of the trivial
solution, provided that it is strong enough: 
\begin{equation}
\sinh \,\Gamma _{1}>\sinh \left( \left( \Gamma _{1}\right) _{\mathrm{cr}
}\right) \equiv \frac{\gamma }{|\sin \,\Gamma _{2}|}\,.  \label{critical}
\end{equation}
Note that the instability is \emph{impossible} in the absence of the local
refractive-index perturbation $\Gamma _{2}$. The instability-onset condition
(\ref{critical}) simplifies in the limiting case when both the loss and gain
parameters are small (while $\Gamma _{2}$ is not necessarily small): 
\begin{equation}
\,\Gamma _{1}>\left( \Gamma _{1}\right) _{\mathrm{cr}}\approx \frac{\gamma 
} {|\sin \,\Gamma _{2}|}\,.  \label{smallcritical}
\end{equation}
Lastly, we notice that the instability is oscillatory: as it follows from
Eq. (\ref{chi}), $\mathrm{Re}\,\chi \neq 0$, unless $\cos \Gamma _{2}=0$.

As the onset of instability in a system combining uniform loss and local
gain is a simple but new issue, for the comparison's sake it is relevant to
briefly consider it in a similar model, viz., the NLS equation: 
\begin{equation}
iu_{t}+\frac{1}{2}Du_{xx}+|u|^{2}u=-i\gamma u+\left( i\Gamma _{1}-\Gamma
_{2}\right) \delta (x)\cdot u,  \label{Schr}
\end{equation}
where $D$ is the spatial-dispersion coefficient, $\gamma $ and $\Gamma
_{1,2} $ having the same meaning as above. If the variable $t$ in Eq. (\ref
{Schr}) is replaced by the propagation distance $z$, and $x$ is realized as
the transverse coordinate in a planar lossy waveguide, Eq. (\ref{Schr}) may
be interpreted as describing spatial solitons in the above-mentioned case
when the gain is applied along a narrow strip in the waveguide. In fact, the
NLS model is a limit case of the BG system for small-amplitude solitons
(see, e.g., Ref. \cite{Litchinitser}); accordingly, Eq. (\ref{Schr}) is a
small-amplitude limit of Eqs. (\ref{o1}) and (\ref{o2}). Nevertheless, it is
pertinent to consider the NLS model parallel to the BG one, as it will help
to understand the results for the BG system.

A perturbation mode in the linearized equation (\ref{Schr}) is sought for as
[cf. Eqs. (\ref{lin1} ) and (\ref{lin2})] 
\begin{equation}
u(x,t)=A_{+}\,\mathrm{exp}(-i\chi t-\kappa x)\,\,\,\mathrm{at}
\;x>0,\,\,u(x,t)=A_{-}\,\mathrm{exp}(-i\chi t+\kappa x)\;\mathrm{at}\;x<0,
\label{Schr_mode}
\end{equation}
with $\mathrm{Re}\,\kappa >0$. The substitution of Eq. (\ref{Schr_mode})
into the linearized version of Eq. (\ref{Schr}) yields 
\begin{equation}
\chi =-\left[ \left( D/2\right) \kappa ^{2}+i\gamma \right] ,
\label{Schr_omega}
\end{equation}
and the integration of Eq. (\ref{Schr}) in an infinitesimal vicinity of $x=0$
leads to $A_{+}=A_{-}$, and 
\begin{equation}
D\kappa =\Gamma _{2}-i\Gamma _{1}.  \label{D}
\end{equation}
Note that the necessary condition $\mathrm{Re\,}\kappa >0$ and Eq. (\ref{D})
show that, in fact, the perturbation mode (\ref{Schr_mode}) exists only in
the case $\Gamma _{2}D>0$. It is easy to understand the meaning of the
latter condition: the inhomogeneity is \textit{attractive} in this case,
hence it can support the local mode.

The substitution of Eq. (\ref{D}) into Eq. (\ref{Schr_omega}) yields a final
result, 
\begin{equation}
\chi =-\left[ \frac{\left( \Gamma _{2}-i\Gamma _{1}\right) ^{2}}{2D}+i\gamma
\right] \,.  \label{Schr_final}
\end{equation}
As it follows from Eq. (\ref{Schr_final}), the instability-onset condition, 
$\mathrm{Im\,}\chi >0$, means, in the present case, 
\begin{equation}
\Gamma _{1}>\left( \Gamma _{1}\right) _{\mathrm{cr}}\equiv D\gamma /\Gamma
_{2}.  \label{Schr_thr}
\end{equation}
Thus, as well as in the case of the hot spot in BG, Eq. (\ref{Schr_thr})
demonstrates that the hot spot in the NLS model cannot give rise to the
instability, unless it contains the imaginary part. Unlike the BG model, the
additional condition $\Gamma _{2}D>0$ is also necessary for the instability.

\subsection{An exact solution in the conservative model}

An exact solution for the pinned soliton is available for the conservative
version of the full nonlinear model, with $\gamma =\Gamma _{1}=0$. In this
case, it is easy to see that Eqs. (\ref{U}) and (\ref{V}) admit an invariant
reduction, $V(x)=-U^{\ast }(x)$ (the asterisk stands for the complex
conjugation), which leads to a single equation, 
\begin{equation}
\left[ i\frac{d}{dx}U+\Gamma _{2}\cos \theta +\delta (x)\right]
U+3|U|^{2}U-U^{\ast }=0.  \label{reduced}
\end{equation}
As it follows from the integration of Eq. (\ref{reduced}) around the point 
$x=0$, the solution must satisfy a boundary condition 
\begin{equation}
U(x=+0)=U(x=-0)\cdot \exp (i\Gamma _{2})\,.  \label{jump}
\end{equation}

An exact soliton-like solution to Eq. (\ref{reduced}), supplemented by the
condition (\ref{jump}), can be found, following the pattern of the exact
solution for the ordinary gap soliton in the model with $\Gamma _{2}=0$ \cite
{Aceves}: 
\begin{equation}
U(x)=\frac{1}{\sqrt{3}}\frac{\sin \theta }{\cosh \left[ \left( x+a\,\mathrm{
\ sgn}\,x\right) \sin \theta -\frac{i}{2}\theta \right] }\,,
\label{solution}
\end{equation}
where $\mathrm{sgn}\,x\equiv \pm 1$ for positive and negative $x$, and the
real parameter $a$ is determined by the relation 
\begin{equation}
\tanh \left( a\sin \theta \right) =\frac{\tan \left( \Gamma _{2}/2\right)} 
{\tan \left( \theta /2\right) }\,.  \label{tan}
\end{equation}
A corollary of the expressions (\ref{solution}) and (\ref{tan}), that will
be used below, is 
\begin{equation}
\left| U(x=0)\right| ^{2}=\frac{2}{3}\left( \cos \Gamma _{2}-\cos \theta
\right) \,.  \label{amplitude}
\end{equation}
From Eq. (\ref{tan}) it follows that the solution exists not in the whole
interval $0<\theta <\pi $, where the ordinary gap solitons are found, but in
a region determined by the constraint that $\left| \tanh \left( a\sin \theta
\right) \right| <1$, i.e., $\left| \tan \left( \Gamma _{2}/2\right) \right|
<\tan \left( \theta /2\right) $, or 
\begin{equation}
\left| \Gamma _{2}\right| <\theta <\pi  \label{interval}
\end{equation}
(which implies that the solutions exist only if $\left| \Gamma _{2}\right|
<\pi $).

Although the exact solutions found above exist for either sign of $\Gamma
_{2}$, it is expected that only in the case $\Gamma _{2}>0$ they may be
stable, as in this case the local inhomogeneity \emph{attracts} the soliton
(which is natural, as positive $\Gamma _{2}$ corresponds to a local
enhancement of the refractive index, and a bright soliton is always
attracted to an optically denser spot) \cite{kdv}. In particular, in the
case of small $\Gamma _{2}$ the soliton may be regarded as a quasiparticle
in an effective inhomogeneity-induced potential 
\begin{equation}
W_{1}(\xi )=-\frac{8\Gamma _{2}}{3}\frac{\sin ^{2}\theta }{\cosh \left( 2\xi
\sin \theta \right) +\cos \theta }\,\,,  \label{W}
\end{equation}
where $\xi $ is a displacement of the soliton's center from the point $x=0$ 
\cite{kdv}. It is obvious that this potential indeed corresponds to the
attraction and repulsion in the cases $\Gamma _{2}>0$ and $\Gamma _{2}<0$.

Note that the exact solution is a single-humped one, with a maximum at the
point $x=0$, if $\Gamma _{2}>0$; in the opposite case, the solution is a
double-humped, with a local minimum at $x=0$ and local maxima at $x=\pm
\left| a\right| $, as in this case Eq. (\ref{tan}) gives $a<0$. In the
limiting case $\theta -\left| \Gamma _{2}\right| \rightarrow 0$ [see Eq. 
(\ref{interval})], Eq. (\ref{amplitude}) shows that $\left| U(x=0)\right|$
vanishes, i.e., the soliton pinned by the attractive inhomogeneity, with 
$\Gamma _{2}>0$, reduces to zero, while the unstable two-humped state pinned
by the repulsive inhomogeneity, with $\Gamma _{2}<0$, goes over into a pair
of two infinitely separated solitons with $\theta =-\Gamma _{2}$.

\subsection{The first-order approximation for the full model}

In the case $\gamma =\Gamma _{1}=0$, Eqs. (\ref{o1}) and (\ref{o2}) conserve
the net energy, 
\begin{equation}
E=\int_{-\infty }^{+\infty }\left[ \left| u(x)\right| ^{2}+\left|
v(x)\right| \right] ^{2}dx\,.  \label{E}
\end{equation}
In the presence of the loss and gain, the exact evolution equation for the
energy is 
\begin{equation}
\frac{dE}{dt}=-2\gamma E+2\Gamma _{1}\left[ \left| u(x)\right| ^{2}+\left|
v(x)\right|^2 \right] |_{x=0}\,.  \label{dE/dt}
\end{equation}
If the coefficients $\gamma $ and $\Gamma _{1}$ are treated as small
perturbations, the balance condition for the energy, $dE/dt=0$, may select a
particular solution, from the family of the exact solutions (\ref{solution})
of the conservative model, which remains, to the first approximation, a
stationary pinned soliton in the full model.

The balance condition following from Eq. (\ref{dE/dt}) demands 
\begin{equation}
\gamma E=\Gamma _{1}\left[ \left| U(x=0)\right| ^{2}+\left| V(x=0)\right|
^{2}\right] \,.  \label{dE/dt=0}
\end{equation}
Substituting, in the first approximation, the unperturbed solution (\ref
{solution}) and (\ref{tan}) into Eq. (\ref{dE/dt=0}), and taking into regard
the definition (\ref{E}), the balance condition can be cast, after some
algebra, in a simple form: 
\begin{equation}
\frac{\theta -\Gamma _{2}}{\cos \Gamma _{2}-\cos \theta }=\frac{\Gamma _{1}}
{\gamma }\,.  \label{complicated}
\end{equation}
As the pinned soliton may only be stable if $\Gamma _{2}>0$, we consider
this case. Note also that, according to Eq. (\ref{interval}), we should
constrain the consideration to the interval $\theta >\Gamma _{2}$, as
otherwise the pinned solitons do not exist in the zero approximation 
($\gamma =\Gamma _{1}=0$).

The pinned soliton selected by Eq. (\ref{complicated}) is expected to
appear, with the increase of the \textit{stress parameter} $\Gamma
_{1}/\gamma $, as a result of some bifurcation. The inspection of Fig. 1,
which displays $(\theta -\Gamma _{2})$ vs. $\Gamma _{1}/\gamma $, as found
from Eq. (\ref{complicated}), shows that the situation is qualitatively
different in the cases $\Gamma _{2}<\pi /2$ and $\Gamma _{2}>\pi /2$.

In the former case, a \textit{tangent} (saddle-node) bifurcation \cite{Iooss}
occurs at a minimum value $\left( \Gamma _{1}/\gamma \right) _{\min }$ of
the stress parameter at which Eq. (\ref{complicated}) has a physical
solution for $\theta $, and two solutions exist for $\Gamma _{1}/\gamma >$ 
$\left( \Gamma _{1}/\gamma \right) _{\min }$. An additional analysis of 
Eq. (\ref{complicated}) demonstrates that, with the variation of $\Gamma _{2}$,
the value $\left( \Gamma _{1}/\gamma \right) _{\min }$ attains an absolute
minimum, $\Gamma _{1}/\gamma =1$, at $\Gamma _{2}=\pi /2$.

With the increase of $\Gamma _{1}/\gamma $, the lower solution branch that
starts at the saddle-node bifurcation point [see Fig. 1(a)] hits the limit
point $\theta =\Gamma _{2}$ [see Eq. (\ref{interval})], where it degenerates
into the zero solution, according to Eq. (\ref{amplitude}). Equation (\ref
{complicated}) shows that this happens at the point $\Gamma _{1}/\gamma
=1/\sin \Gamma _{2}$. On the other hand, it was shown above [see Eq. (\ref
{smallcritical})] that, precisely at the same point, the zero solution
becomes unstable, in the limit of small $\Gamma _{1}$ and $\gamma $.
According to the general stability-exchange principle \cite{Iooss}, the fact
that the zero-solution branch gets unstable after its collision with another
solution branch implies that the latter branch was already unstable.
Therefore, we conclude that the branch which appears at the saddle-node
bifurcation and ceases to exist hitting the zero solution, is an unstable
saddle.

The other (upper) branch generated by the saddle-node bifurcation [Fig.
1(a)] continues until it attains the maximum value $\theta =\pi $ relevant
to the physical solutions, which happens at 
\begin{equation}
\frac{\Gamma _{1}}{\gamma }=\left( \frac{\Gamma _{1}}{\gamma }\right) _{\max
}\equiv \frac{\pi -\Gamma _{2}}{1+\cos \Gamma _{2}}\,.  \label{max}
\end{equation}
This branch corresponds to the node-type solution which appears at the
saddle-node bifurcation point, therefore it has a chance to be stable.
However, it may be unstable against perturbations that are not taken into
regard by this elementary consideration (for instance, the possibility of a
shift of the soliton's center from the point $x=0$ was not taken into
regard). The actual situation with the stability of pinned solitons in the
model including the loss and gain is rather complicated, see the next
section.

In the case $\Gamma _{2}>\pi /2$, the situation is different, as the
saddle-node bifurcation is imaginable in this case, occurring in the
unphysical region $\theta <\Gamma _{2}$, see Fig. 1(b). The only physical
branch of the solutions appears at the point $\Gamma _{1}/\gamma =1/\sin
\Gamma _{2}$, where it crosses the zero solution, lending it the
instability, which the branch presumably had in the unphysical region. The
stability-exchange principle which was already mentioned above suggests
that, from the viewpoint of the present analysis, this branch becomes stable
when it crosses into the physical region, $\theta >\Gamma _{2}$. However, as
well as the other branch considered above for the case $\Gamma _{2}<\pi /2$,
the present one may be subject to instabilities of other types. This branch
ceases to be a physical one at the point (\ref{max}).

At the border between the two generic cases considered above, i.e., at 
$\Gamma _{2}=\pi /2$, the saddle-node bifurcation occurs exactly at the point 
$\theta =\pi /2$, see Fig. 1(c). In this case, the destabilization of the
zero solution happens at the same point.

The situation is different in the case $\Gamma _{2}=0$ [see Fig. 1(d)], when
the hot spot has no refractive-index-perturbation component, and Eq. (\ref
{complicated}) takes the form 
\begin{equation}
\frac{\theta }{2\sin ^{2}\left( \theta /2\right) }=\frac{\Gamma _{1}}{\gamma 
}\,.  \label{noGamma2}
\end{equation}
In this case, as it was stressed above, the zero solution is never
destabilized, in accordance with which the solution branches do not cross
the axis $\theta =0$ in Fig. 1(d). The lower branch, which asymptotically
approaches the $\theta =0$ axis, must be unstable (this is a generic feature
in the case when the amplitude of the solution decreases with the increase
of the stress parameter \cite{Iooss}), hence the upper branch may be stable
within the framework of the present analysis. However, direct numerical
simulations presented below demonstrate that, in the case $\Gamma _{2}=0$,
the pinned soliton is always unstable against the displacement of its center
from the point $x=0$.

It may be relevant to compare these results with those that can be obtained
for pinned solitons in the NLS model containing the loss and ``hot spot'',
see Eq. (\ref{Schr}). NLS solitons may only exist if $D>0$, therefore we
adopt the normalization $D=1$ in Eq. (\ref{Schr}), and it makes sense to
consider only the case when the inhomogeneity is attractive, i.e., $\Gamma
_{2}>0$, otherwise the pinned soliton has no chance to be stable. Note that,
once we choose $D>0$ and $\Gamma _{2}>0$, the zero solution may be unstable,
according to the results presented above.

In the conservative limit, $\gamma =\Gamma _{1}=0$, the pinned NLS soliton
is given by a commonly known solution, 
\begin{eqnarray}
u &=&\eta \,\mathrm{sech}\left( \eta \left( |x|+a\right) \right) \,\exp
\left( \frac{i}{2}\eta ^{2}t\right) ,  \label{pinned Schr} \\
a &=&\frac{1}{2\eta }\ln \left( \frac{\eta +\Gamma _{2}}{\eta -\Gamma _{2}}
\right) \,,  \label{a Schr}
\end{eqnarray}
where $\eta $ is an intrinsic parameter of the solution family. Note that,
as we assume $\Gamma _{2}>0$, Eq. (\ref{a Schr}) yields $a>0$, hence the
expression (\ref{pinned Schr}) has a single maximum at $x=0$.

If now $\gamma $ and $\Gamma _{2}$ are introduced as small parameters, the
energy-balance condition for this solution can be easily cast in the form 
$\eta =-\Gamma _{2}+2\gamma /\Gamma _{1}$. In view of the relation (\ref{a
Schr}), the actual solution exists in the region $\eta >\Gamma _{2}$, or,
eventually, in the interval 
\begin{equation}
0<\Gamma _{1}<\gamma /\Gamma _{2}\,.  \label{gGG}
\end{equation}

Comparing this result with the expression (\ref{Schr_thr}) that determines
the instability threshold for the zero solution, and taking into regard that 
$D=1$, we conclude that the pinned soliton singled out by the balance
condition disappears [crosses into the unphysical region, cf. Fig. 1(a)],
with the increase of the stress parameter, $\Gamma _{1}/\gamma $, exactly at
the point where the zero solution loses its stability. According to the
above-mentioned stability exchange principle, this implies that the zero
solution inherits its instability from the soliton, hence the soliton
solution, given by Eqs. (\ref{pinned Schr}) and (\ref{a Schr}), is
definitely \textit{unstable} in all the region of its existence.

This conclusion demonstrates that the above results for the pinned solitons
in the BG model are nontrivial, as they give the pinned gap soliton a chance
to be stable, which is not possible at all in the simpler NLS model. Actual
stability of the pinned gap solitons will be studied below by means of
direct simulations.

\section{Numerical results for pinned solitons}

\subsection{The approximation for the delta-function}

For the simulations, we have to adopt a numerical form of the $\delta 
$-function in Eqs.~(\ref{o1}),(\ref{o2}) and (\ref{U}), (\ref{V}). We use the
same numerical scheme as in the recent work \cite{kdv}. The scheme
discretizes the coordinate $x$ by $501$ grid points $x_{j}$, 
$j=-250,...,-1,0,+1,...+250$. As an approximation to the $\delta $-function,
the following function is defined on a set of $2N+1$ grid points in the
central part of the integration domain, located symmetrically around zero, 
\begin{equation}
\widetilde{\delta }\left( x_{n-\left( N+1\right) }\right) \equiv 
\begin{array}{ll}
A\,\mathrm{cos}\,\left( \frac{n-\left( N+1\right) }{2N+1}\pi \right)  & 
\mathrm{for}\;n=1,\ldots ,2N+1, \\ 
0 & \mathrm{elsewhere.}
\end{array}
\label{approxd}
\end{equation}
\noindent The normalization factor $A$ is defined so as to maintain the
canonical normalization of the $\delta $-function, $\int_{-\infty }^{+\infty
}\widetilde{\delta }(x)dx\equiv \sum_{j}\widetilde{\delta }\left(
x_{j}\right) \Delta x=1$, which yields 
\begin{equation}
A=\left[ \Delta x\sum_{n=1}^{2N+1}\mathrm{cos}\,\left( \frac{n-\left(
N+1\right) }{2N+1}\pi \right) \right] ^{-1},  \label{A}
\end{equation}
$\Delta x$ being the spacing of the grid (in fact, $\Delta x=0.04)$. In most
cases presented below, we use $N=2$ [then Eq.~(\ref{A}) with $\Delta x=0.04$
yields $A=\left[ \left( \allowbreak 1+\sqrt{5}\right) \Delta x\right]
^{-1}\approx \allowbreak 7.\,\allowbreak 726$], which makes the 
$\delta$-function quite narrow indeed.

\subsection{Stability of the pinned solitons in the conservative model}

Since exact stationary solutions to Eqs.~(\ref{U}) and (\ref{V}) for the
pinned soliton are available in the case $\gamma =\Gamma _{1}=0$, in the
form of Eq.~(\ref{solution}) supplemented by Eq.~(\ref{tan}), numerical test
of their stability is straightforward. We simulated the stability by means
of the split-step method applied to Eqs.~(\ref{o1}) and (\ref{o2}),
employing the fast Fourier transform. The exact solution (\ref{solution})
was taken as the initial configuration, and the corresponding value $\theta
_{\mathrm{in}}$ of $\theta $ was varied. The values $\theta _{\mathrm{in}
}<\left| \Gamma _{2}\right| $, at which the exact solution does not exist
[see Eq. (\ref{interval})] were probed too. In this case, Eq. (\ref{tan})
yields an imaginary value of $a$, and the initial configuration was taken in
the form of Eq. (\ref{solution}) with the imaginary $a$. Even though the
latter configuration is not a stationary solution, it is still nonsingular
and localized, so it can be used to launch the PDE simulation.

As expected from what was mentioned above, in the case $\Gamma _{2}<0$ all
the pinned states of the solitons are found to be unstable. Solitons are
pushed away from the point $x=0$, in accord with the expectation that the
inhomogeneity is repulsive. It was also observed that, as $\left| \Gamma
_{2}\right| $ increases, at $\Gamma _{2}$ $\leq $ $-0.7$ a small soliton is
left behind around the point $x=0$ after the main pulse has separated from
it; however, the residual soliton is also unstable and gradually decays into
radiation.

For positive $\Gamma _{2}$, typical results regarding the stability of the
pinned solitons are displayed in Fig.~2. A conclusion is that there is a 
\emph{single value} $\theta _{\mathrm{stab}}\approx \pi /2$ of the soliton
parameter $\theta $, such that if $\theta _{\mathrm{in}}<\theta _{\mathrm{\
stab}}$, the soliton decays into radiation, as is seen in Fig.~2(a).
Solitons with $\theta _{\mathrm{in}}>\theta _{\mathrm{stab}}$ relax into a
stable one with $\theta =\theta _{\mathrm{stab}}$, see Fig.~2(b). Finally, a
soliton with $\theta _{\mathrm{in}}=\theta _{\mathrm{stab}}$ directly gives
rise to the stable soliton, see Fig. 2(c).

The examples shown in Fig.~2 pertain to $\Gamma _{2}=0.4$, and similar
results were obtained for other values of $\Gamma _{2}$. Available
computational power imposes a limitation on accuracy with which $\theta _{
\mathrm{stab}}$ can be identified. However, it was found that, for $\Gamma
_{2}=0.1$, the decrease of the soliton's amplitude, which is defined as 
$|u(x=0)|$, is less than $1\%$ after the evolution time $T=200\pi $, if 
$\theta _{\mathrm{in}}$ is taken from the interval $\left( 0.49\pi <\theta _{
\mathrm{in}}<0.52\pi \right) $, hence, in any case, $\theta _{\mathrm{stab}
}\left( \Gamma _{2}=0.1\right) $ belongs to the same interval. For a much
larger value of the perturbation parameter, $\Gamma _{2}=1.1$, the
corresponding interval is $0.51\pi <\theta _{\mathrm{in}}<0.55\pi $, hence 
$\theta _{\mathrm{stab}}\left( \Gamma _{2}=1.1\right) $ belongs to this
region. Generally, $\theta _{\mathrm{stab}}$ slightly increases with $\Gamma
_{2}$.

Figure~3 summarizes these results in the form of a plot in the $(\Gamma
_{2},\theta _{\mathrm{in}})$ plane, which shows the regions where the
initial soliton relaxes to the stable one or decays into radiation. In the
region $\theta _{\mathrm{in}}<$ $\left| \Gamma _{2}\right| $, where the
initial configurations are not true stationary solutions, this configuration
decays into radiation immediately.

These results, obtained for the conservative model with the local
inhomogeneity of the refractive index, are very similar to those reported in
Ref. \cite{kdv} for the stability of the solitons pinned by an attractive
inhomogeneity in the form of a local suppression of the Bragg grating. A
noticeable common feature of the results is the existence of the \emph{single}
(up to the numerical accuracy available) value $\theta _{\mathrm{stab}
}\approx \pi /2$ of the parameter $\theta $ which the established soliton
may assume. In both conservative models (the ones considered here and in
Ref. \cite{kdv}), $\theta _{\mathrm{in}}$ relaxes to $\theta _{\mathrm{stab}
} $ if $\theta _{\mathrm{in}}>\theta _{\mathrm{stab}}$, and the soliton
decays into radiation if $\theta _{\mathrm{in}}<\theta _{\mathrm{stab}}$,
i.e., the soliton with $\theta =\theta _{\mathrm{stab}}$ may be called a 
\textit{semi-attractor}. In fact, it strongly resembles \textit{semi-stable}
solitons, which are stable against small perturbations in the linear
approximation, but may be unstable if terms quadratic in the perturbations
are take into regard. Semi-stable solitons were recently studied in another
context, as the so-called embedded solitons, see Ref. \cite{emb} and
references therein.

The fact that all the solitons with $\theta _{\mathrm{in}}>\theta _{\mathrm{
stab}}$, where $\theta _{\mathrm{stab}}$ is slightly larger than $\pi /2$,
relax to the value $\theta =\theta _{\mathrm{stab}}$, may be related to a
known property of the ordinary solitons in the unperturbed BG model ($\gamma
=\Gamma _{1}=\Gamma _{2}=0$): they are unstable if $\theta >\theta _{\mathrm{
\ cr}}^{(0)}\approx 1.011\cdot \left( \pi /2\right) $ \cite{Barash}. Thus,
at least in the case when $\Gamma _{2}$ is small, it is natural to expect
that any pinned soliton with $\theta >\pi /2$ will relax, as a result of the
instability, to a value close to $\theta _{\mathrm{cr}}^{(0)}$. What is less
obvious, is the decay of the solitons with 
$\theta <\theta _{\mathrm{stab}}$, and the fact that 
$\theta _{\mathrm{stab}}$ so weakly depends on $\Gamma_{2}$ (see Fig. 3).

\subsection{The pinned soliton in the lossy medium with the localized gain}

In direct simulations of the full model, which includes the loss and local
gain, the exact solution (\ref{solution}) of the conservative version was
again used as the starting point. The evolution of the solution was
simulated at a fixed value of the loss parameter $\gamma $. The local gain 
$\Gamma _{1}$ was varied in order to determine its value(s) at which the
soliton settles down to a stable pinned soliton.

Figure 4 shows the evolution of the soliton's amplitude, defined as $\left|
u(x=0)\right| $, vs. $t$, when the localized gain $\Gamma _{1}$ is varied.
The other parameters are fixed, so that 
\begin{equation}
\Gamma _{2}=0.5,\gamma =0.0316,\,\mathrm{\,}\theta _{\mathrm{in}}=0.5\pi .
\label{fixed}
\end{equation}
For a small value of $\Gamma _{1}$ ($\Gamma _{1}=0.04208$ in Fig.~4), which
is insufficient to balance the loss, the soliton decays. For a slightly
larger $\Gamma _{1}=0.04209$, the soliton's amplitude grows, then it
temporarily settles down (at the value $1.47$ in Fig. 4), and, eventually,
regular oscillations set in. A long simulation, up to $t=600\pi $ (see Fig.
4) shows that the intrinsic vibrations of the soliton are completely stable.
The waveforms $\left| u(x,t)\right| $ and $\left| v(x,t)\right| $, obtained
at the end of the simulation for $\Gamma _{1}=0.04209$, are shown in Figure
5(a).

When the gain $\Gamma _{1}$ is further increased, the initial growth of the
soliton's amplitude is sharper; however, it is found that it again
temporarily settles down to a nearly constant value close to the same level
of $1.47$ as above, which is followed by the onset of persistent
oscillations. When $\Gamma _{1}$ is still larger, the eventual oscillatory
state becomes chaotic with large fluctuations. The corresponding waveforms
of $\left| u(x)\right| $ and $\left| v(x)\right| $ at the end of the
simulation ($t=300\pi $) for $\Gamma _{1}=0.057$ are shown in Fig. 5(b). It
can be seen that conspicuous radiation tails are attached to the soliton,
which implies a permanent energy leakage from it. This extra loss adds up to
the direct dissipative loss, both being compensated by the localized gain.
If $\Gamma _{1}$ is too large, the radiation wave field outside the main
pulse grows to such an extent that the resulting waveform cannot be regarded
as a localized one. In fact, in this case it is observed that the main pulse
separates from the point $x=0$, drifts away, and dies down due to the loss.
However, the strong localized gain generates a new ``soliton'' around $x=0$,
which later drifts away again, this process repeating itself
quasi-periodically.

An important feature of these results is that a stable (even though it is
vibrating) soliton is possible not at a single value of the gain, that
exactly compensates the loss, but in a finite interval of values of the
gain. The energy balance is maintained, in this case, through permanent
emission of radiation by the soliton, which compensates the excessive gain.
It is relevant to mention that a very similar mechanism, which gives rise to
stable \textit{nonequilibrium solitons} in an overpumped system of a
different type (that, however, also originates in nonlinear optics -- the
so-called split-step model), was recently considered in detail in Ref. \cite
{Radik}. In that case too, the stability of the soliton is provided by the
emission of radiation that balances the excess gain.

A further insight in sustained intrinsic vibrations of the pinned soliton,
and the transition from the regular oscillations to dynamical chaos, is
provided by consideration of the spectrum of the function $\left|
u(x=0,t)\right| $. In the established oscillatory regime, the spectrum was
computed at several different values of $\Gamma _{1}$, while the other
parameters were kept constant as per Eq. (\ref{fixed}). Figure 6(a) shows
the spectrum for $\Gamma _{1}=0.04209$, which is the value barely enough to
compensate the loss. It can be seen that the established oscillations are
quasi-harmonic, with a single well-pronounced frequency $2.9$ (in arbitrary
units), and an additional tiny spectral component at the frequency $\approx
2 $ (which is, apparently, incommensurate with the main one).

Figure 6(b) shows the spectrum for $\Gamma _{1}=0.5633$, which is similar to
that in Fig.~6(a). The main frequency shifts down to a value about $2.8$,
with two other visible components found at the frequencies $0.8$ and $1.4$.
Then, suddenly, at a slightly larger gain, $\Gamma _{1}=0.5634$, many new
conspicuous spectral components emerge, which is shown in Fig.~6(c), and
corresponds to (apparently) chaotic intrinsic vibrations of the established
soliton. The same behavior is observed at $\Gamma _{1}=0.5635$. At $\Gamma
_{1}=0.5636$, the picture suffers another abrupt change [see Fig. 6(d)]: the
power spectrum reverts back to the simple three-frequency-component
structure reminiscent of the situation at lower $\Gamma _{1}$, cf.
Fig.~6(b). A transition from a chaotic behavior, (presumably) accounted for
by a strange attractor, to a simple quasi-harmonic behavior is known in the
theory of dynamical systems, where it is called a ``boundary crisis'' of the
chaotic attractor \cite{Hilborn}.

The picture revealed by the simulations changes the third time at $\Gamma
_{1}=0.5640$, with reappearance of a many-component chaotic-like spectrum,
similar to that in Fig.~6(c). The chaotic behavior continues to higher
values of $\Gamma _{1}$. Figure~6(e) shows the spectrum at $\Gamma
_{1}=0.05690$, where its structure is not simply a multi-component one, but
continuous, which is characteristic for well-developed dynamical chaos.

Another way to describe basic properties of the pinned solitons in the full
model is to identify, for various initial values of $\theta _{\mathrm{in}}$,
a minimum value $\left( \Gamma _{1}\right) _{\min }$ of the gain which is
necessary to overcome the loss. Figure~7 shows the evolution of the
soliton's amplitudes as a function of time for $\theta _{\mathrm{in}}=0.2\pi
,\,0.5\pi ,\,$ and $0.9\pi $, the corresponding minimum values being $\left(
\Gamma _{1}\right) _{\min }=0.0239$, $0.0133$, and $0.0131$, while the other
parameters are fixed, $\gamma =0.01$ and $\Gamma _{2}=0.5$ [note that all
these values of $\theta _{\mathrm{in}}$ exceed $\Gamma _{2}$, hence the
corresponding exact solitons in the conservative model do exist, according
to Eq. (\ref{interval})]. Thus, unlike the conservative model, in the full
model, values of $\theta _{\mathrm{in}}$ essentially smaller than $\pi /2$
may give rise to a stable pinned soliton (with intrinsic vibrations).
However, the smaller the difference $\theta _{\mathrm{in}}-\Gamma _{2}$, the
larger value of $\Gamma _{1}$ is necessary, as, according to Eqs. (\ref
{dE/dt}) and (\ref{amplitude}), the rate at which the localized gain
supplies energy to the soliton decreases $\sim \left( \theta _{\mathrm{in}
}-\Gamma _{2}\right) $ as $\theta _{\mathrm{in}}-\Gamma _{2}\rightarrow 0$.
On the other hand, analysis of the simulation results shows that the
characteristics of the established soliton do not depend on the initial
value of $\theta _{\mathrm{in}}$ which excited it, but solely on the values
of $\gamma $ and $\Gamma _{1,2}$, i.e., the established soliton is a genuine 
\emph{\ attractor}.

Then, effects caused by varying the loss parameter $\gamma $ were
investigated. Because of the necessity to satisfy the energy balance
condition, $\Gamma _{1}$ needs to be changed to track the variation of 
$\gamma $. For each value of $\gamma $, we tried to find the minimum size of 
$\Gamma _{1}$ that supports a stable soliton. Results of these numerical
experiments, obtained for fixed $\Gamma _{2}=0.5$ and $\theta _{\mathrm{in}
}=\pi /2$, and a set of values $\gamma
=0.000316,\,0.001,0.00316,\,0.01,\,0.0316,\,$and $0.1$, are displayed in
Fig. 8. The respective minimum-gain values were found to be $\left( \Gamma
_{1}\right) _{\min }=0.000422,\,0.00140,\,0.00422,\,0.0133,\,0.04209$, and 
$0.1327$. It is interesting to note that, except for the second case, when
the ratio $\left( \Gamma _{1}\right) _{\min }/\gamma $ is $1.40$, in all the
other ones the ratio takes values between $1.32$ and $1.34$.

It is clearly seen from Fig. 8 that the amplitude of the established soliton
monotonically increases with the growth of $\gamma $ (which is accompanied
by the growth of the minimum gain necessary to support the soliton). It is
also seen that it was never possible to produce a truly stationary soliton,
but in some cases ($\gamma =0.000316,\,0.00316,\,0.01,$ and $0.1$) it was
possible to generate nearly stationary solitons with a small amplitude of
intrinsic vibrations. In other cases ($\gamma =0.001$ and $0.0316$), varying 
$\Gamma _{1}$ by steps as small as it was admitted by the numerical scheme,
it was \emph{not} possible to adjust the gain so that to suppress the
internal vibrations, i.e., the established soliton remained a breather,
rather than anything close to a fixed-point state.

In all the cases presented in Fig. 8, $\gamma $ and $\Gamma _{1}$ are small
enough to treat them as perturbations. Then, if the established soliton
assumes a nearly stationary form, it is natural to expect that it must be
close to the solution (\ref{solution}) found in the conservative model, with
some value of $\theta $, and this $\theta $ must be related to $\gamma $ and 
$\Gamma _{1}$ as per Eq. (\ref{complicated}).

It was checked that the quasi-stationary solitons, in those cases in when
they were found, are indeed close to the wave form (\ref{solution}). The
corresponding values of $\theta $ were identified by means of the
least-squared-error fit to the expression~(\ref{solution}). Then, for thus
found values of $\theta $ and given values of $\gamma $, the equilibrium
values of the gain $\Gamma _{1}$ were calculated as predicted by the
analytical formula (\ref{complicated}). Results of this are presented in
Table 1.

A noticeable fact obvious from Table 1 is that, in all the cases, the
numerically found equilibrium value of the gain exceeds the analytically
predicted one by $9$ to $14$ per cent. Because in all the cases, as it was
stressed above, the established solitons are not completely stationary, a
natural conjecture is that the slightly vibrating soliton continuously emits
energy at a low rate, and this extra energy loss makes it necessary to have
the gain somewhat larger than that which compensates the direct dissipative
loss as per Eq. (\ref{complicated}).

As concerns the comparison of the full model with its conservative
counterpart, we recall that, in the conservative model, the stable pinned
soliton always assumes a single value of $\theta $ for given $\Gamma _{2}$
(and this value very weakly depends on $\Gamma _{2}$, always being slightly
larger than $\pi /2$, see Fig. 3). On the contrary to this, in the full
model the quasi-stationary soliton may be stable in a range of the values of 
$\theta $, as it is evident from Table 1.

Finally, it has also been checked whether stable pinned solitons can be
found when the ``hot point'' does not perturb the refractive index, i.e., 
$\Gamma _{2}=0$. As a result, it has been concluded that any finite positive 
$\Gamma _{2}$ (the smallest value tried was $\Gamma _{2}=0.01$) may support a
stable soliton in the pinned state, but if $\Gamma _{2}=0$, the pulse set at 
$x=0$ finally drifts away, and then decays due to the loss. An explanation
to this finding may be that all the solitons found in the model with loss
and gain emit some radiation, see above, and asymmetric fluctuations in the
emission rate create a weak random force that drives the soliton away.

\subsection{Soliton pinned between two repulsive inhomogeneities}

The soliton may be stably pinned not only by an attractive inhomogeneity,
but also between two separated repulsive ones (in the present context, each
one will represent a locally suppressed refractive index, corresponding to 
$\Gamma _{2}<0$). The consideration of this configuration is interesting by
itself, and it also may be used to design a soliton-based optical
oscillator, in which the eigenfrequency is easily controlled by the choice
of the separation between the two repulsive points. In particular, in the
framework of the perturbation theory 
(for small $\left| \Gamma _{2}\right|$), the soliton may be regarded 
as a quasiparticle in the external potential \begin{equation}
W_{2}(\xi )=W_{1}\left( \frac{1}{2}L-\xi \right) +W_{1}\left( \frac{1}{2}
L+\xi \right) ,  \label{W2}
\end{equation}
where the potential $W_{1}(\xi )$ is given by Eq. (\ref{W}) (with $\Gamma
_{2}<0$), and $L$ is the separation between the two defects.

We simulated the dynamics of this configuration in some detail, but only for
the conservative case, $\gamma =\Gamma _{1}=0$. First of all, if $L$ is
smaller than the proper size of the soliton, it sees the pair of the
inhomogeneities, in the first approximation, as a single repulsive center,
hence stable bound states are not possible. Within the framework of the
perturbation theory, Eqs. (\ref{W2}) and (\ref{W}) make it possible to
predict a critical value, $\left( \Delta \xi \right) _{\mathrm{cr}}$, at
which a stable equilibrium appears for the first time at $x=0$. The
corresponding expression is cumbersome, but it easy to verify that $\left(
\Delta \xi \right) _{\mathrm{cr}}$ monotonically decreases, with $\theta $
varying from $0$ to $\pi /2$, from $\left( \Delta \xi \right) _{\mathrm{cr}
}=\infty $ to the minimum value $\left( \Delta \xi \right) _{\mathrm{cr}
}=\ln \left( \sqrt{2}+1\right) \approx 0.88$.

Direct simulations at finite $\Gamma _{2}$ demonstrate [see an example in
Fig. 9] that, in the case of relatively small $L$, when the pinned state of
the soliton is unstable, the dynamical evolution does not trivially reduce
to pushing the soliton aside; instead, a generic outcome is \emph{splitting}
of the soliton in two, which is accompanied by a spontaneous symmetry
breaking (in some cases, for instance if $\theta _{\mathrm{in}}=0.7\pi $,
the other parameters being the same as in Fig. 9, the instability develops
so slowly that it was not possible to identify the outcome).

With the increase of $L$, stabilization of the soliton trapped between the
repulsive inhomogeneities becomes possible. The trapped states seem most
stable around the value $L=3.84$, see an example in Fig. 10. In this case,
systematic simulations reveal a feature which strongly resembles the one
reported above for the single attractive inhomogeneity in the conservative
model: an established trapped state is stable for a single (up to the
accuracy of numerical simulations) value of $\theta $, which is very close
to $\pi /2$; if $\theta _{\mathrm{in}}>\pi /2$, the soliton sheds off some
radiation and eventually relaxes to the said single value of $\theta $ (see
Fig. 10), while if $\theta _{\mathrm{in}}<\pi /2$, the soliton gradually
decays into radiation. Thus, the single-valuedness of the stable soliton in
the conservative model appears to be a generic property. For still larger
values of $L$, the pinned state is less robust; in particular, a soliton
with $\theta _{\mathrm{\ in}}>\pi /2$ may split, instead of relaxing to the
stable one with $\theta \approx \pi /2$.

\section{Collision of a moving soliton with the localized gain}

Once the existence of stable pinned soliton has been established, the next
natural step is to consider a possibility of capturing a free moving soliton
by the ``hot spot''. To this end, the soliton was first generated far from
the spot by means of the Newton-Raphson method, as a stationary solution in
the reference frame moving at some velocity $c$; a range of the velocities 
$0\leq c\leq 0.7$ was thus investigated. Then, the collision was considered,
running direct simulations of Eqs.~(\ref{o1}) and (\ref{o2}).

First, the collision experiment was performed in the conservative model,
with $\Gamma _{1}=\gamma =0$. The parameter plane $(c,\Gamma _{2})$ was
explored with $c$ taking values $0.1,\,0.2,\,\ldots ,0.7$, and $\Gamma _{2}$
taking values $0.1,\,0.2,\,\ldots ,0.9$, while $\theta _{\mathrm{in}}$ was
kept constant at $0.7\pi $.

If the inhomogeneity is weak, the moving soliton passes through it, see an
example in Fig.~11(a). When the inhomogeneity strength is larger, $\Gamma
_{2}\,\,_{\sim }^{>}\,0.5$, a part of the soliton still passes through it,
but another part of the soliton's energy is captured by the local defect to
form a pinned soliton, an example of which is shown in Fig.~11(b). Some
radiation bouncing in the backward direction can also be observed when 
$\Gamma _{2}$ is large, or when the incident soliton is fast. Naturally, more
energy is trapped by the defect if $\Gamma _{2}$ is larger [Fig. 11(c)], and
less energy is trapped if the soliton is faster. However, the value $\Gamma
_{2}\approx 0.5$, at which the trapping begins, only weakly depends on the
soliton's velocity $c$. Figure~12 summarizes these results, showing a border
in the $(c,\Gamma _{2})$ plane between the passage and partial-capture
regions.

Next, we consider the collisions in the full model, with $\gamma =0.01$ and 
$\Gamma _{1}$ $=0.015$. Results reported in the previous section show that a
stable pinned soliton exists at these values of the loss and gain (the
collisions were simulated only for small values of $\gamma $, as otherwise
the soliton will be strongly attenuated still before the collision). The
initial value of the soliton's parameter was again $\theta _{\mathrm{in}
}=0.7\pi $.

The results obtained for the full model are not very different from those
for the conservative one. When the gain $\Gamma _{2}$ is small, the soliton
passes through, and if $\Gamma _{2}$ is larger, a part of the energy is
trapped to form a pinned soliton. A difference from the conservative model
is that the value of $\Gamma _{2}$ at which the inhomogeneity starts to
capture a part of the soliton's energy in the conservative model is
approximately independent of its velocity: $\Gamma _{2}\approx 0.5$ if 
$c>0.1 $, while in the lossy model, this value of $\Gamma _{2}$ increases
with $c$, as is seen in Fig.~12.

Another representative set of numerical data can be displayed for a fixed
value of the soliton's velocity, $c=0.1$, while the parameter $\theta_{
\mathrm{in}}$ of the incident soliton takes values $0.1\pi ,\,0.2\pi
,\,\ldots ,\,0.9\pi $, and $\Gamma _{2}=0.1,\,0.2,\,\ldots ,\,0.9$. These
results are presented here only for the conservative model.

Simulations show that, if the moving soliton is heavy (large $\theta_{
\mathrm{in}}$) or the inhomogeneity is weak, the soliton passes it. Heavier
solitons can pass through a stronger defect. If the inhomogeneity is strong 
($\Gamma _{2}$ is large), the soliton gets trapped, which is always
accompanied by emission of radiation in both the forward and backward
directions, and the radiation can further self-trap into secondary solitons.
At small $\Gamma _{2}$, little energy is scattered away in either direction.
If $\Gamma _{2}$ is larger, more energy is scattered forward, and when 
$\Gamma _{2}$ is still larger ($\Gamma _{2}\approx 0.9$), more energy is
scattered backwards, cf. Fig.~11(c). Figure 13 summarizes the results
obtained for the interaction of the moving soliton and the localized
attractive defect in the conservative model for the fixed velocity, $c=0.1$

It is natural to compare the results obtained for the conservative model
with those reported in Ref. \cite{kdv} for the interaction of the moving gap
soliton with an attractive inhomogeneity in the form of a local suppression
of the Bragg grating. In that case, when the soliton was heavy (large 
$\theta $), the interaction effectively reverted from attraction to
repulsion, so that the incident soliton could bounce back. In the present
model, this unusual behavior has never been observed.

\section{Conclusion}

In this work, we have introduced a model of a lossy nonlinear fiber grating
with a ``hot spot'' combining the localized gain and attractive
inhomogeneity of the refractive index; the spot can be created by means of
doping a short segment of the fiber. In the absence of the loss and gain, a
family of exact solutions for pinned solitons was found. In the full model
including loss and gain, the instability threshold for the zero solution was
found; it was concluded that the instability is not possible without the
presence of the imaginary part of the local gain, i.e., a localized
perturbation of the refractive index. Further, for small values of the loss
and gain, it was predicted what soliton is selected by the energy-balance
condition. Parallel to this, it was shown that, in the simpler model based
on the NLS equation, the pinned soliton can never be stable in the presence
of the loss and local gain.

In direct simulations, we have found \ that a single pinned soliton is
stable in the conservative fiber-grating model. It is a semi-attractor:
solitons with a larger energy relax to it via emission of radiation, while
the ones with smaller energy completely decay into radiation. The same
conclusion is obtained for solitons trapped between two repulsive
inhomogeneities. In the full model with the loss and gain, all the stable
pinned pulses demonstrate persistent internal vibrations and emission of
radiation. Sometimes, they are almost stationary solitons, and in these
cases the prediction based on the energy balance underestimates the
necessary gain by $9\%$ to $14\%$, which is explained by the extra radiation
loss. If the loss and gain increase, the intrinsic vibrations become chaotic.

Collisions of free moving solitons with the ``hot spot'' were simulated too.
The passage and capture regimes were identified for the solitons in the
conservative and full models; the capture is only partial, which actually
implies splitting of the soliton. If was also found that, if a large part of
the soliton's energy is radiated away, it may self-trap into secondary
solitons.

\section*{Acknowledgement}

B.A.M. appreciates hospitality of the Optoelectronic Research Centre at the
City University of Hong Kong.

\newpage

\section*{Tables}

\begin{tabular}[t]{|l|l|l|l|l|}
\hline
$\gamma $ & $\theta $ & $\left( \Gamma _{1}\right) _{\mathrm{num}}$ & 
$\left( \Gamma _{1}\right) _{\mathrm{anal}}$ & $\frac{\left( \Gamma
_{1}\right) _{\mathrm{num}}-\left( \Gamma _{1}\right) _{\mathrm{anal}}} 
{\left( \Gamma _{1}\right) _{\mathrm{anal}}}$ \\ \hline
$0.000316$ & $0.5\pi$ & $0.000422$ & $0.000386$ & $0.0944$ \\ \hline
$0.00316$ & $0.595\pi$ & $0.0042$ & $0.00369$ & $0.1373$ \\ \hline
$0.01$ & $0.608\pi$ & $0.01333$ & $0.01165$ & $0.1442$ \\ \hline
$0.1$ & $0.826\pi$ & $0.1327$ & $0.121$ & $0.0967$ \\ \hline
\end{tabular}

\bigskip

Table 1. Values of the loss parameter $\gamma $ at which quasi-stationary
stable pinned solitons were found by the adjustment of the gain $\Gamma _{1}$
(see Fig. 8), while the refractive-index perturbation is fixed, $\Gamma
_{2}=0.5$. The corresponding values of the gain, $\left( \Gamma _{1}\right)
_{\mathrm{num}}$, are also included, together with the values of the soliton
parameter $\theta $ which provide for the best fit of the quasi-stationary
solitons to the analytical solution (\ref{solution}). The values $\left(
\Gamma _{1}\right) _{\mathrm{anal}}$ are those predicted, for given $\gamma $
and $\theta $, by the energy-balance equation (\ref{complicated}), which
does not take the radiation loss into account.

\newpage

\section*{Figure captions}

Fig. 1. Analytically predicted solution branches for the pinned soliton in
the case of weak loss and gain. Shown is $\protect\theta -\Gamma_2$ vs. the
stress parameter $\Gamma_1/\protect\gamma$. (a) An example of the case 
$\Gamma_2 < \protect\pi/2$ is displayed with $\Gamma_2 = \protect\pi/4$; (b)
an example of the case $\Gamma_2 > \protect\pi/2$ is displayed with 
$\Gamma_2 = 3\protect\pi/4$; (c) $\Gamma_2 = \protect\pi/2$; 
(d) $\Gamma_2= 0$. In the last case, nontrivial solutions appear at 
the point $\protect\theta
=0.7442\protect\pi,\, \Gamma_1/\protect\gamma = 1.3801$, and at large values
of $\Gamma_1/\protect\gamma$ the continuous curve asymptotically approaches
the horizontal axis. In all the panels, the dashed lines show a formal
continuation of the solutions in the unphysical regions, $\protect\theta <
\Gamma_2$, and $\protect\theta > \protect\pi$. In the panels (a), (b), and
(c), the trivial solution, $\protect\theta - \Gamma_2 =0$, is shown by the
solid line where it is stable; in the case corresponding to the panel (d),
all the axis $\protect\theta =0$ corresponds to the stable trivial solution.
(Note that all quantities plotted are dimensionless.)

Fig. 2. Typical results illustrating the stability and instability of
pinned solitons in the case $\protect\gamma=\Gamma_1 =0$, and 
$\Gamma_2=0.4$. Each panel shows the evolution of $|u(x,t)|$, 
starting with the exact
pinned-soliton configuration (the evolution of $|v(x,t)|$ is quite similar.)
(a) $\protect\theta_{\mathrm{in}} = 0.4\protect\pi$ is smaller than $\protect
\theta_{\mathrm{stab}}$: the soliton decays into radiation.
(b) $\protect\theta_{\mathrm{in}} =0.8\protect\pi > \protect\theta_{
\mathrm{stab}}$: the initial soliton transforms itself into a stable one,
shedding off excess energy in the form of radiation.
(c) $\protect\theta_{\mathrm{in}} =0.5\protect\pi \approx \protect
\theta_{\mathrm{stab}}$. Direct appearance of the stable soliton.

Fig. 3. A summary of results obtained for the stability of pinned
solitons, plotted in the $(\Gamma_2,\protect\theta_{\mathrm{in}})$ plane, in
the conservative model with $\protect\gamma=\Gamma_1=0$. In the upper
region, where $\protect\theta > \protect\theta_{\mathrm{stab}}$, initial
solitons evolve into the stable one, shedding off extra energy. In the lower
region, where $\protect\theta < \protect\theta_{\mathrm{stab}}$, solitons
completely decay into radiation. Beneath the lower solid line, which borders
the region where $\left| \Gamma_2 \right| < \protect\theta < \protect\pi$,
see Eq.~(\ref{interval}), stationary solutions for the pinned solitons do
not exist. Accordingly, an initial pulse taken as a formal ``soliton", with
an imaginary root of Eq. (\ref{tan}) substituted for $a$ (see the text), is
immediatley destroyed.

Fig. 4. Evolution of the amplitude of the pinned soliton in the full model
with loss and gain, in the case with $\Gamma_2 = 0.5$, $\protect\gamma
=0.0316 $ and $\protect\theta_{\mathrm{in}}=0.5 \protect\pi$. If 
$\Gamma_1=0.04208$, the gain is insufficient to balance the loss, and the
soliton decays. When $\Gamma_1 = 0.04209$, the soliton initially grows, and
its intrinsic vibrations develop. When $\Gamma_1$ takes a slightly larger
value, $0.04215$, the initial growth of the amplitude is steeper, which
again results in the establishment of regular intrinsic vibrations (in this
case, the oscillations are very similar to those supported by $\Gamma_1 =
0.04209$). When $\Gamma_1$ is essentially larger, for instance, $\Gamma_1
=0.057$, interal vibrations of the pinned soliton become chaotic (the latter
case shown by the dotted line).

Fig. 5. The profiles of $\left| u(x,t) \right|$ (solid lines) and $\left|
v(x,t) \right|$ (dashed lines) at the end of the simulation (narrow peaks
placed at $x=0$ mark the ``hot spot"). The values of $\protect\gamma$, 
$\Gamma_2$, and $\protect\theta_{\mathrm{in}}$ are the same as in Fig. 4. (a) 
$\Gamma_1=0.04209$ is barely enough to support the soliton against the loss.
In this case, the soliton emits radiation at a low rate. 
(b) $\Gamma_1=0.057$. The soliton emits radiation at a high rate.

Fig. 6. The frequency spectrum of the time-dependent amplitude $|u(x=0,t)|$
of the pinned soliton at different values of the local gain $\Gamma_1$,
after persistent vibrations set it: (a) $\Gamma_1=0.04209$, (b) 
$\Gamma_1=0.05633$, (c) $\Gamma_1=0.05634$, (d) $\Gamma_1=0.5636$, and (e) 
$\Gamma_1=0.05690$. In all the cases, $\protect\gamma=0.0316$, $\Gamma_2= 
0.5$, and $\protect\theta_{\mathrm{in}}=0.5\protect\pi$.

Fig. 7. The soliton's amplitude $\left| u(x=0,t)\right|$ vs. $t$ for three
different values of the initial amplitude $\protect\theta_{\mathrm{\ in}}$.
In each case, the value of the gain $\Gamma_1$ is chosen as the minimum one
which can support the establishment of a soliton. Other parameters are
fixed: $\Gamma_2 = 0.5$, and $\protect\gamma=0.01$.

Fig. 8. The amplitude of the soliton, $\left|u(x=0,t) \right|$, vs. $t$ for
different values of the loss parameter $\protect\gamma$. Each time, the
value of the gain $\Gamma_1$ is chosen as the smallest one which leads to
the establishment of the soliton. Other parameters are fixed: $\Gamma_2 =
0.5$, and $\protect\theta_{\text{in}}=0.5\protect\pi$.

Fig. 9. The interaction of a soliton with a pair of repulsive points 
($\Gamma_2=-0.7$), with a relatively small separation between them, $L=1.84$,
in the conservative model. (a) The initial configuration, with $\protect
\theta_{\mathrm{in}}=\protect\pi/2$. (b) The result of the interaction:
splitting of the soliton into two pulses, which is accompanied by a
spontaneous symmetry breaking. The solid lines show $|u|$, and the dashed
lines show $|v|$. Note that, in the initial configuration, $|u|$ and $|v|$
are indiscernible.

Fig. 10. Stable soliton captured between the repulsive points ($\Gamma_2 =
-0.5$) with the separation $L = 3.84$ between them, in the conservative
model. Shown is the evolution of the field $|u(x,t)|$. The initial value of
the soliton parameter is $\protect\theta_{\mathrm{in}} =0.7 \protect\pi$.

Fig. 11. Collision of a moving soliton, with fixed values $\protect\theta =
0.7\protect\pi$ and $c=0.4$, and the inhomogeneity in the conservative model
(the inhomogeneity is shown by a narrow peak which, for an unessential
reason, is shifted from the point $x=0$). The lower and upper panels show,
respectively, the evolution of the field $|u(x,t)|$ in terms of the contour
plots, and the waveforms $|u(x)|$ and $|v(x)|$ (solid and dashed lines) at
the end of the simulation (note that the $u$- and $v$-components are
asymmetric in the moving solitons). (a) The soliton passes through a weak
defect with $\Gamma_2=0.2$.
(b) A stronger defect, with $\Gamma_2=0.5$, captures a part of the
energy of the passing soliton, to form a small-amplitude pinned one. Another
small part of the energy bounces back in the form of radiation.
(c) If the defect is still stronger, $\Gamma_2=0.9$, the shares of
the trapped and bounced energy are larger.

Fig. 12. Borders in the parametric plane ($c,\Gamma_2 $) between regions in
which the moving soliton with fixed $\protect\theta =0.7\protect\pi$ passes
the defect or gets partially trapped. The solid line is the border in the
conservative model, with $\protect\gamma=\Gamma_1=0$. The dashed line is the
border in the full model with $\protect\gamma=0.01$ and $\Gamma_1=0.015$.

Fig. 13. Regions in the parametric plane ($\protect\theta,\,\Gamma_2$) of
the conservative model in which the moving soliton with fixed $c=0.1$ passes
the defect or gets partially trapped.

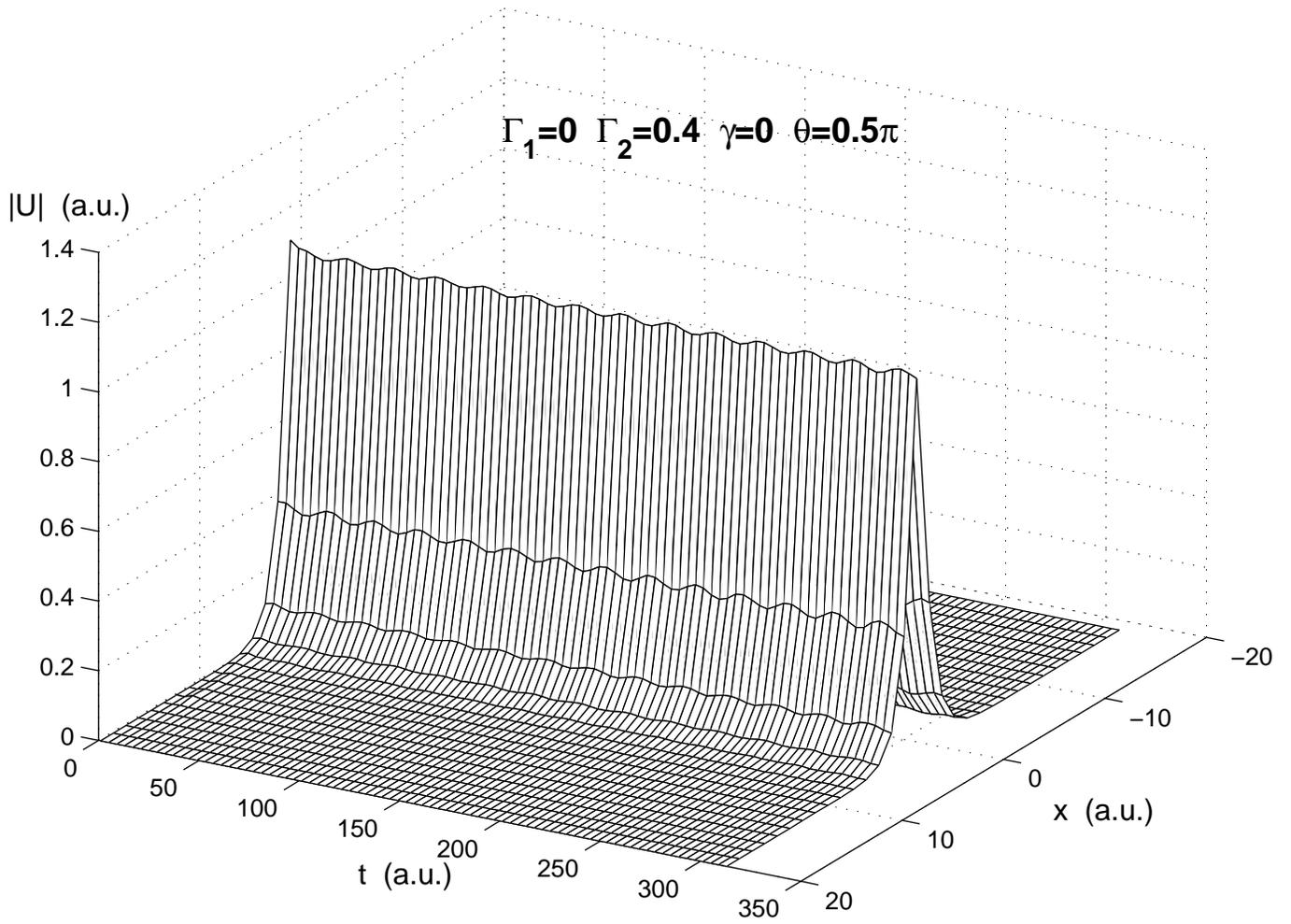
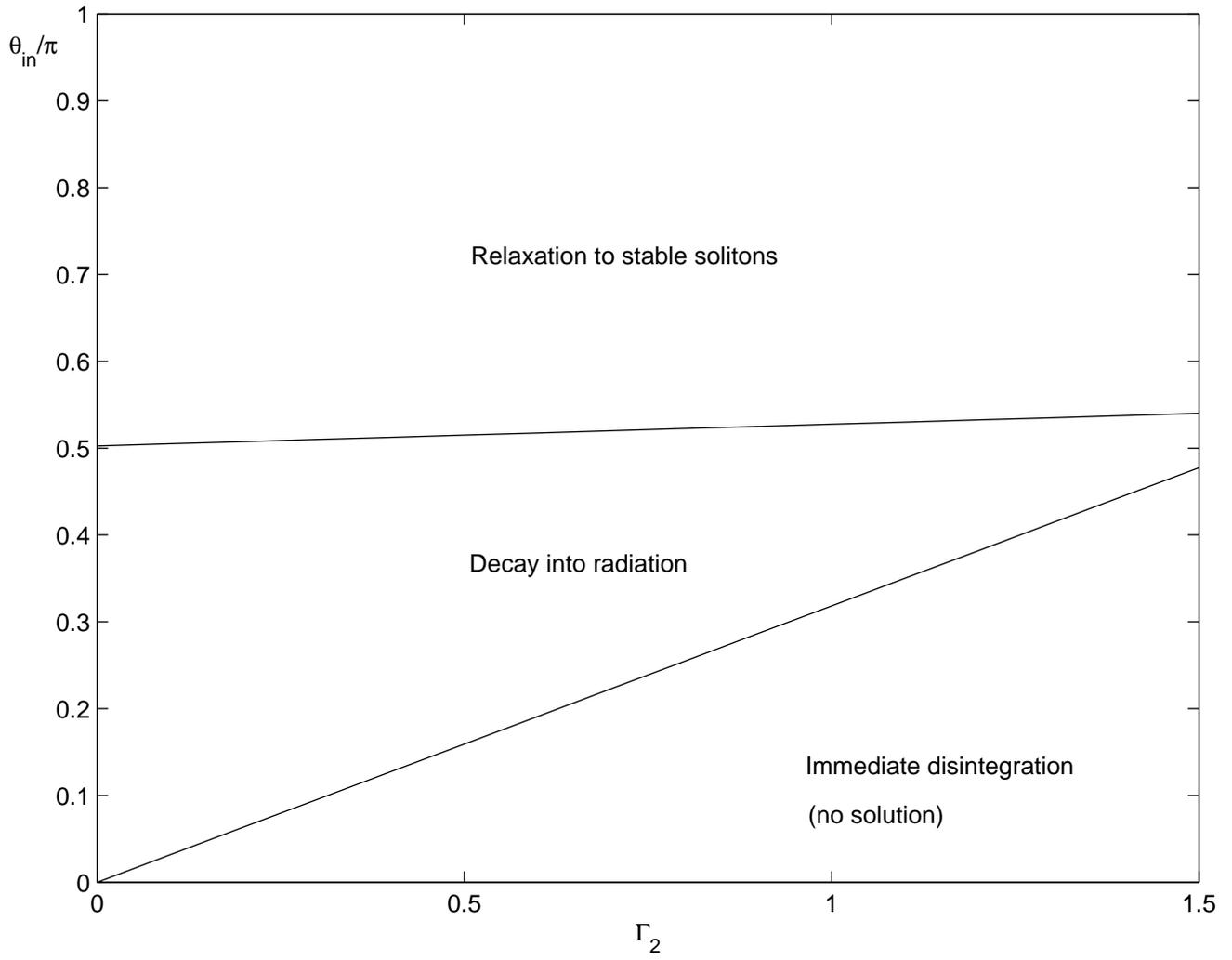
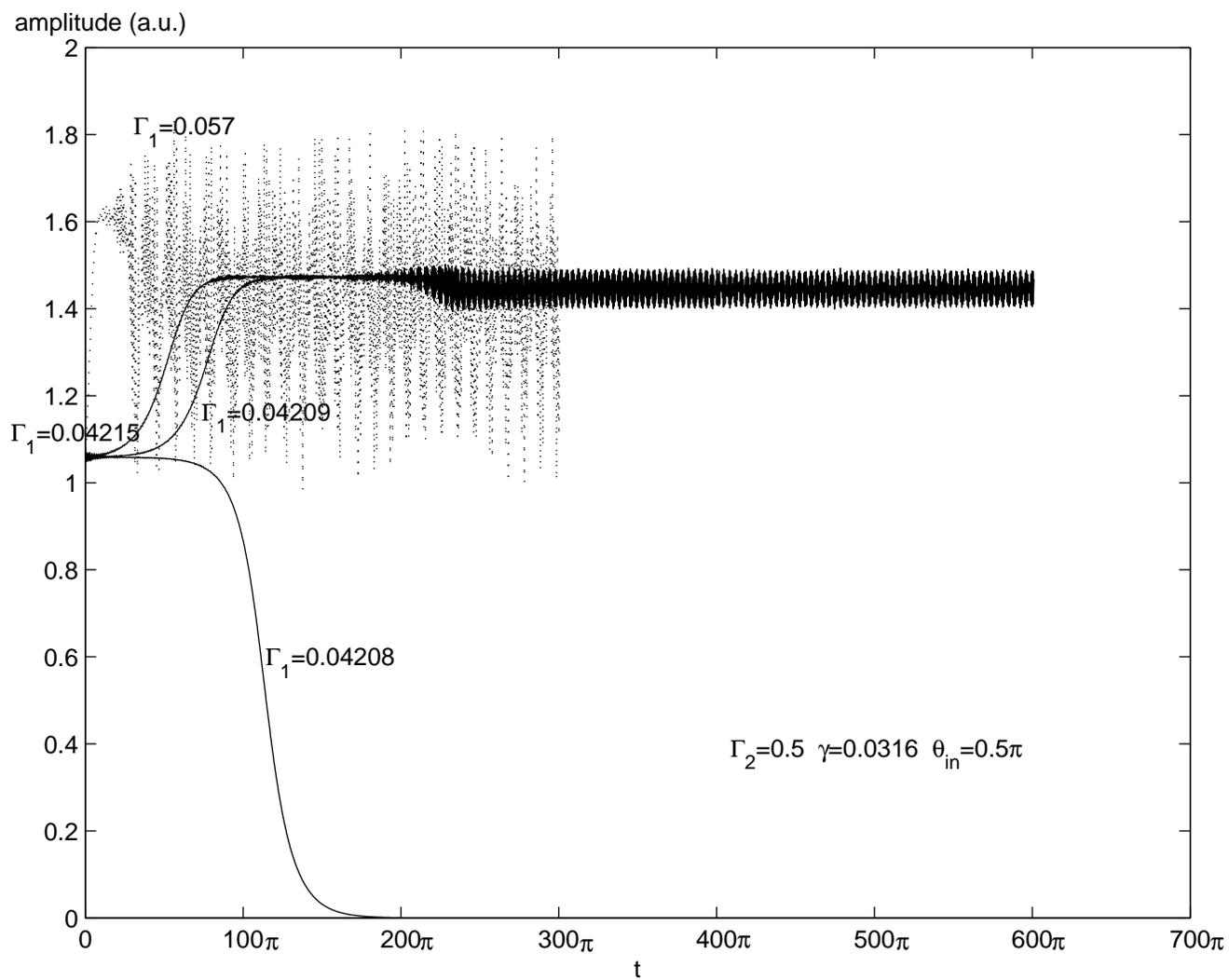
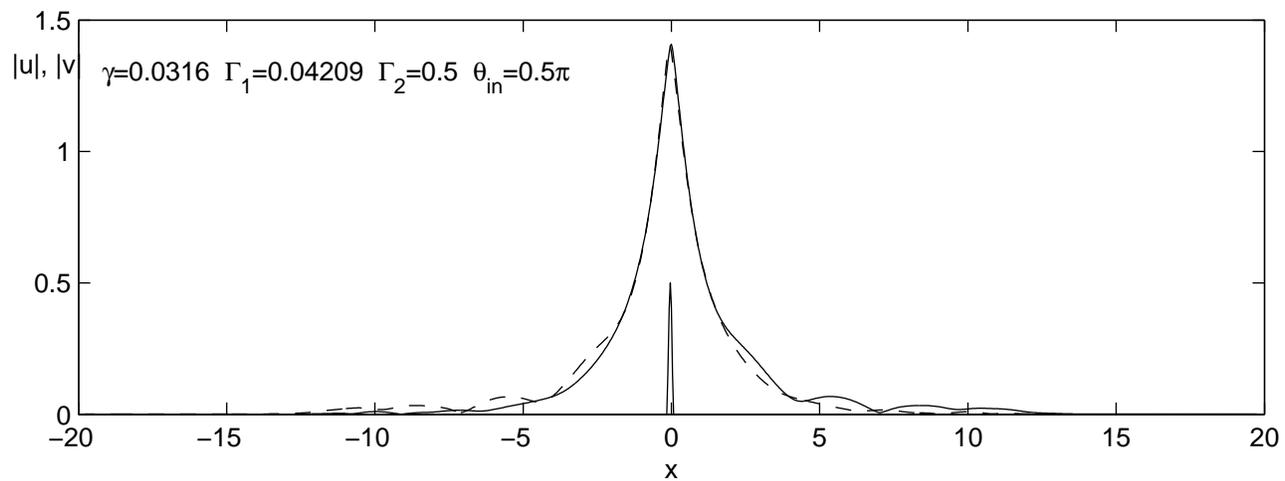
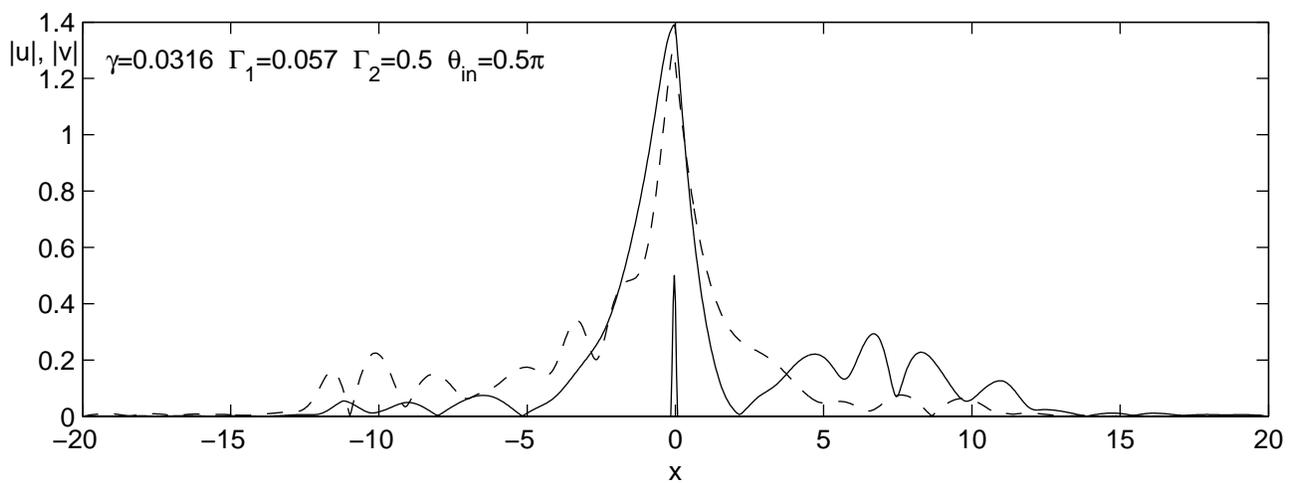
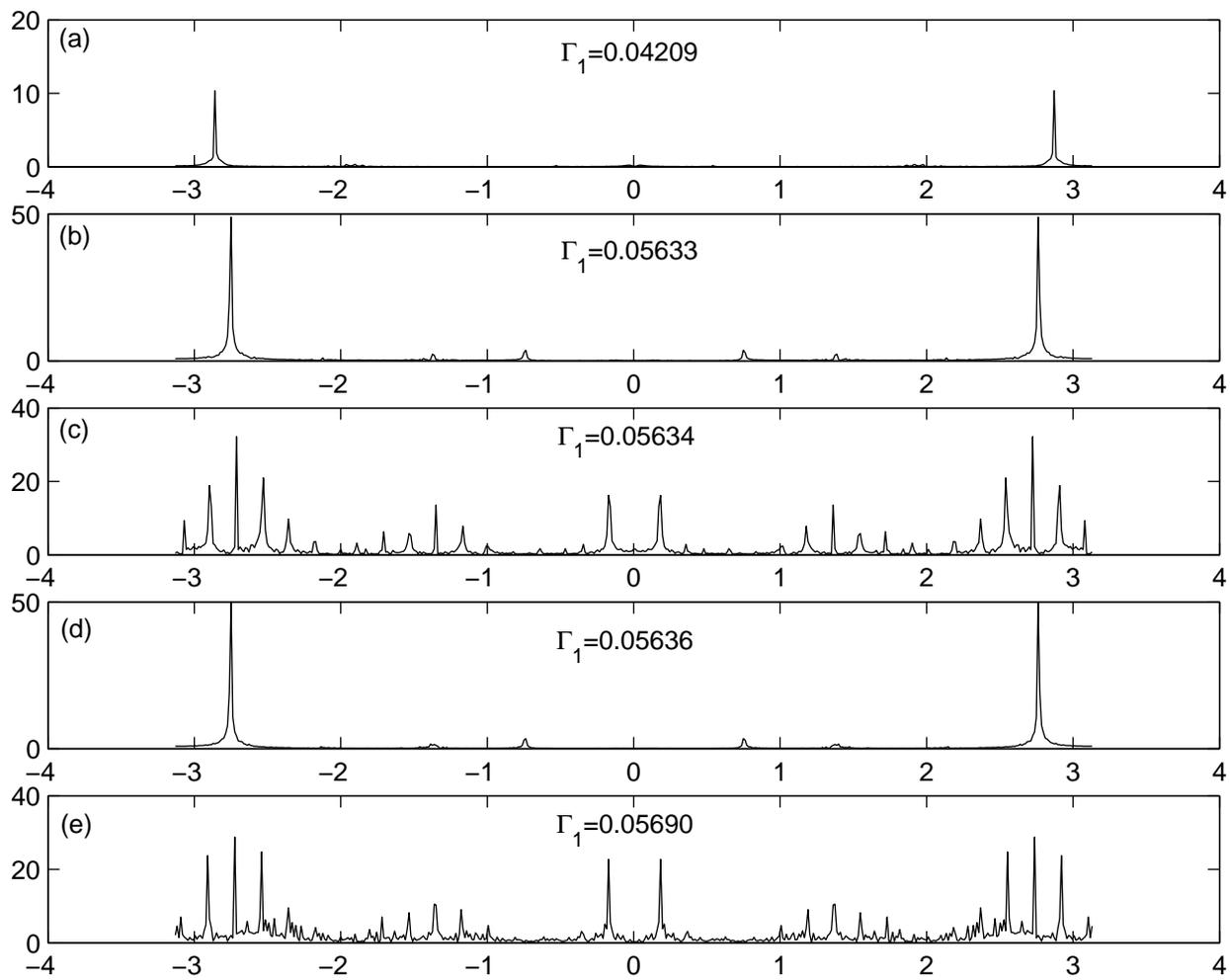
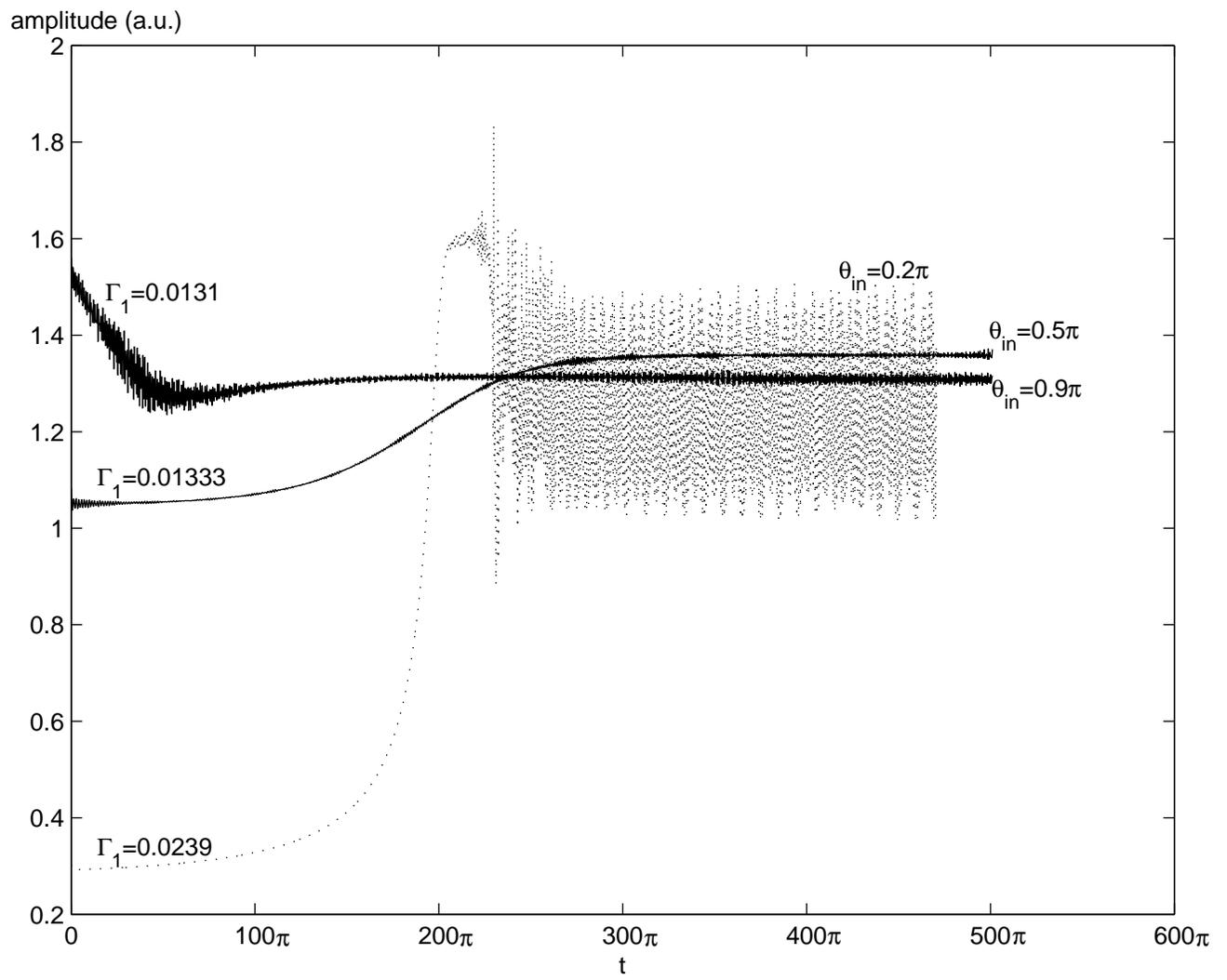
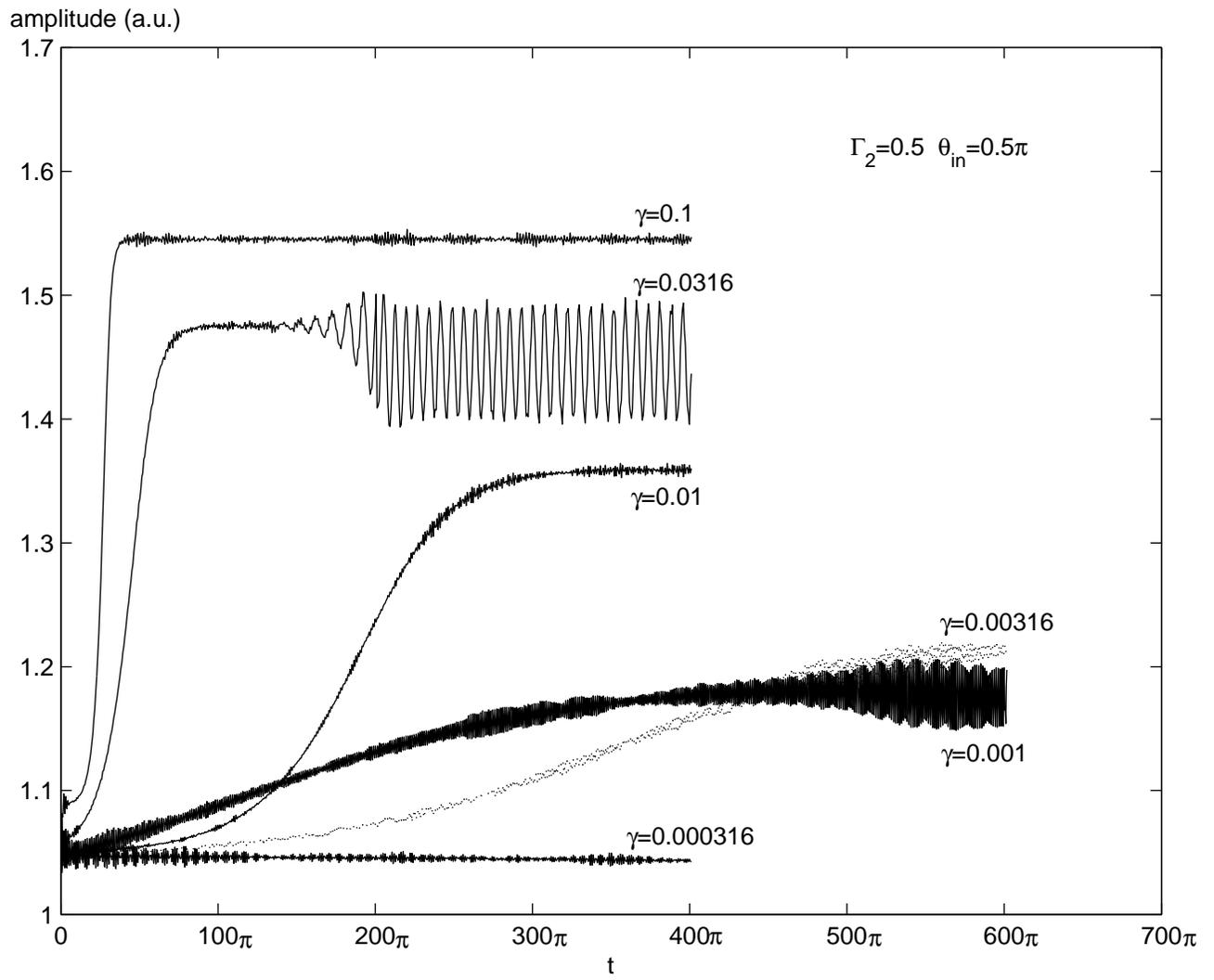
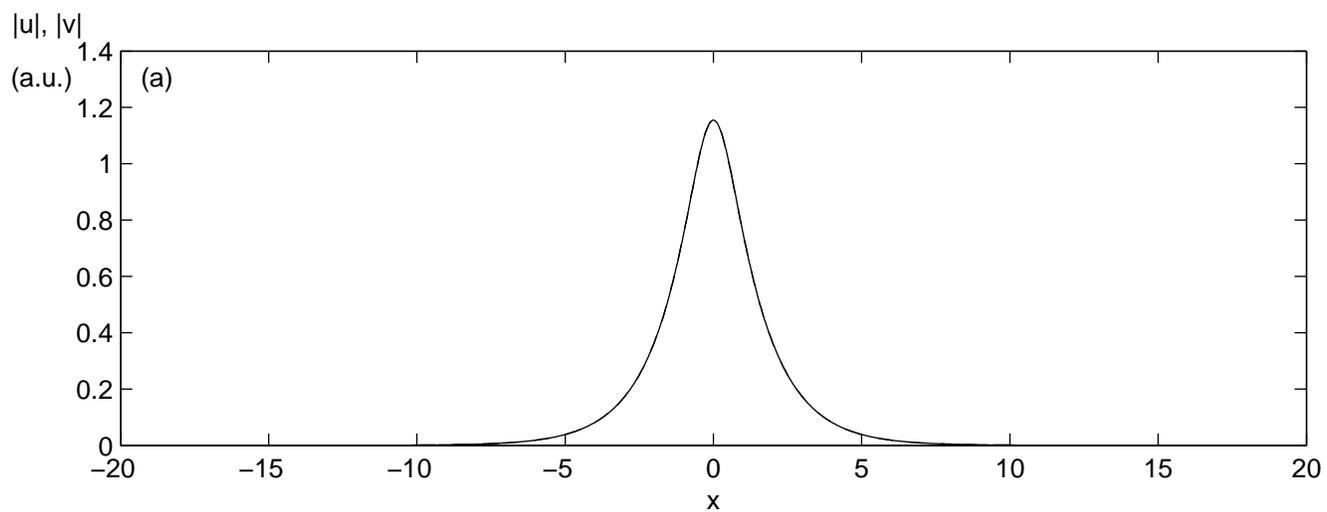
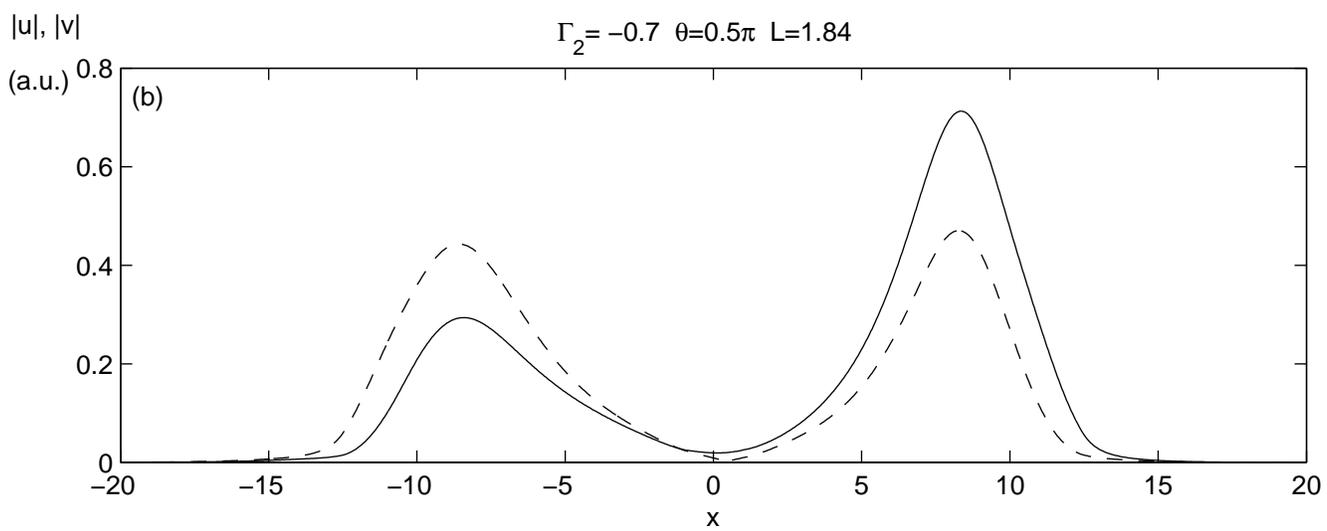
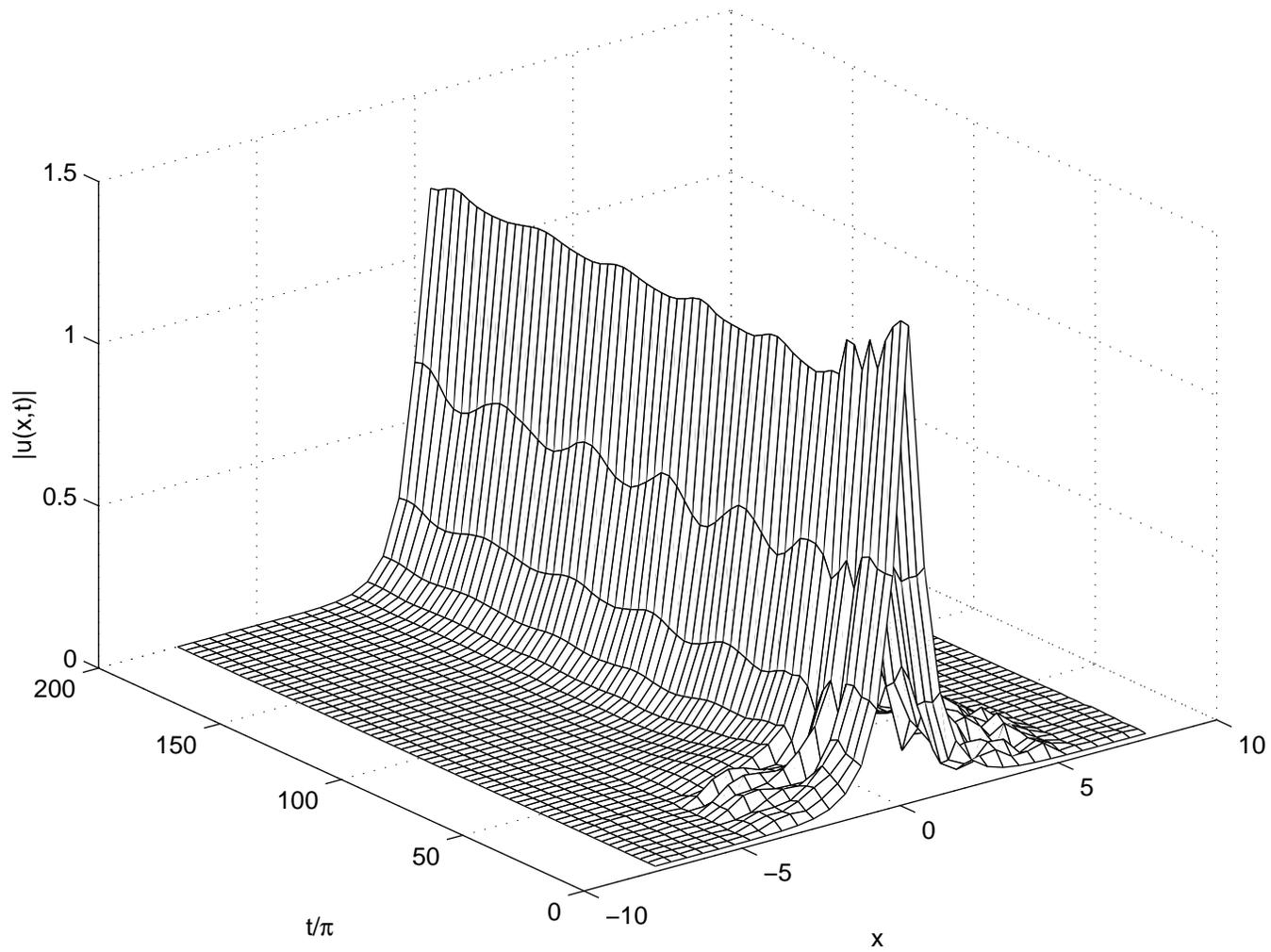
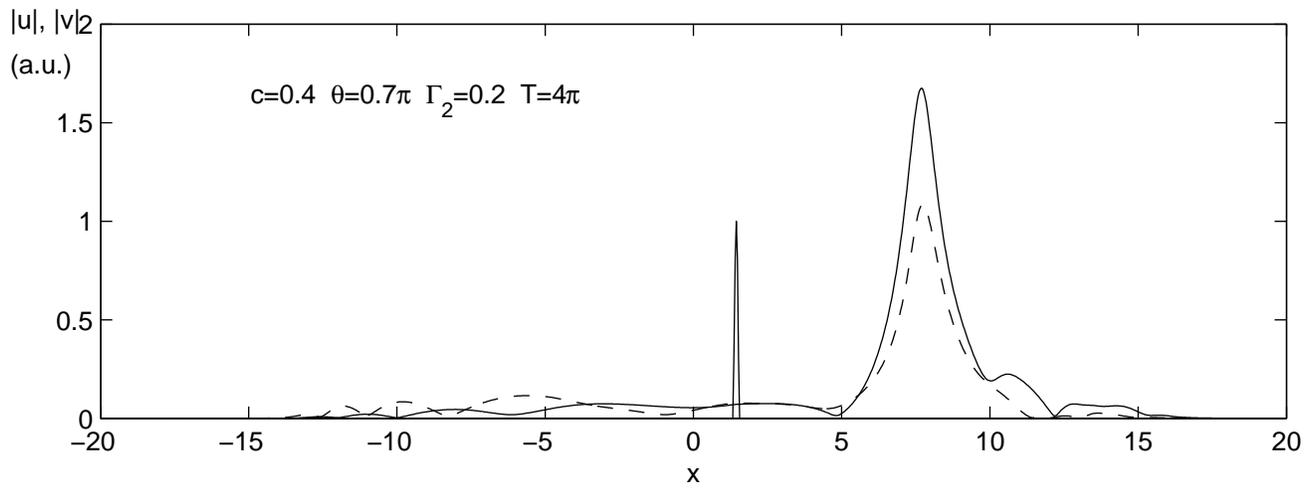
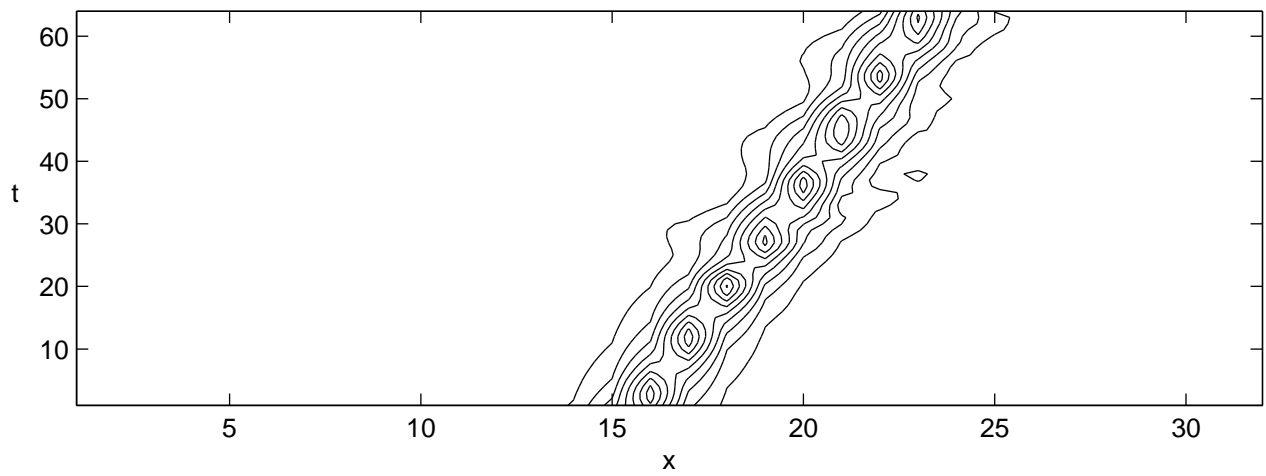
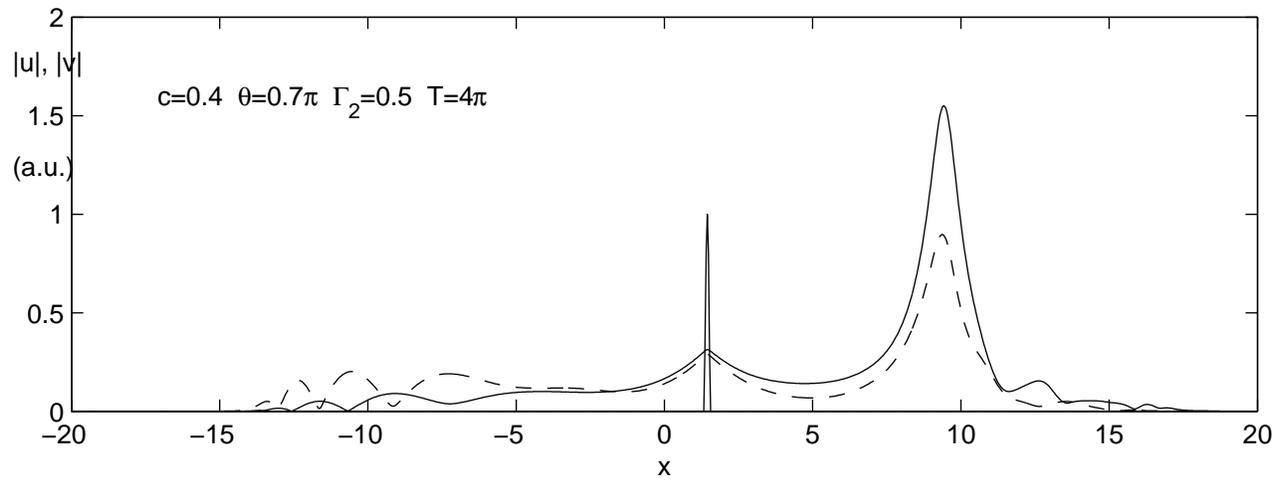
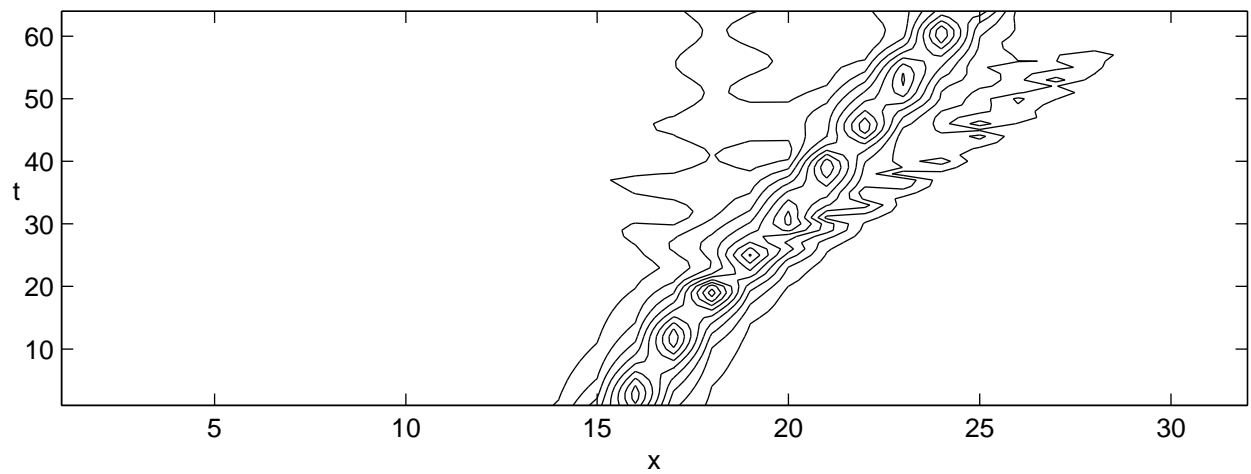
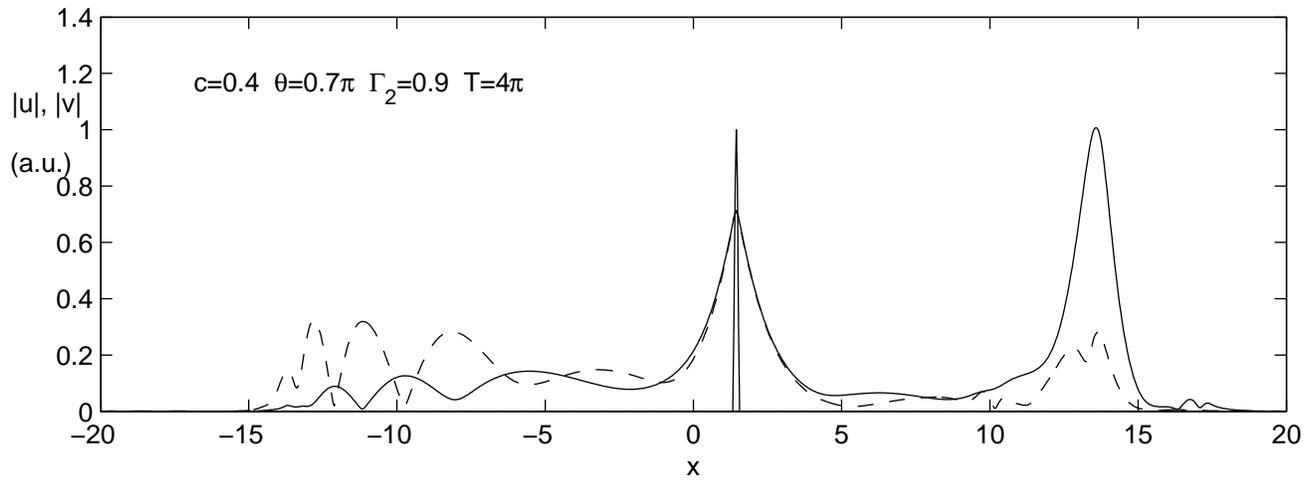
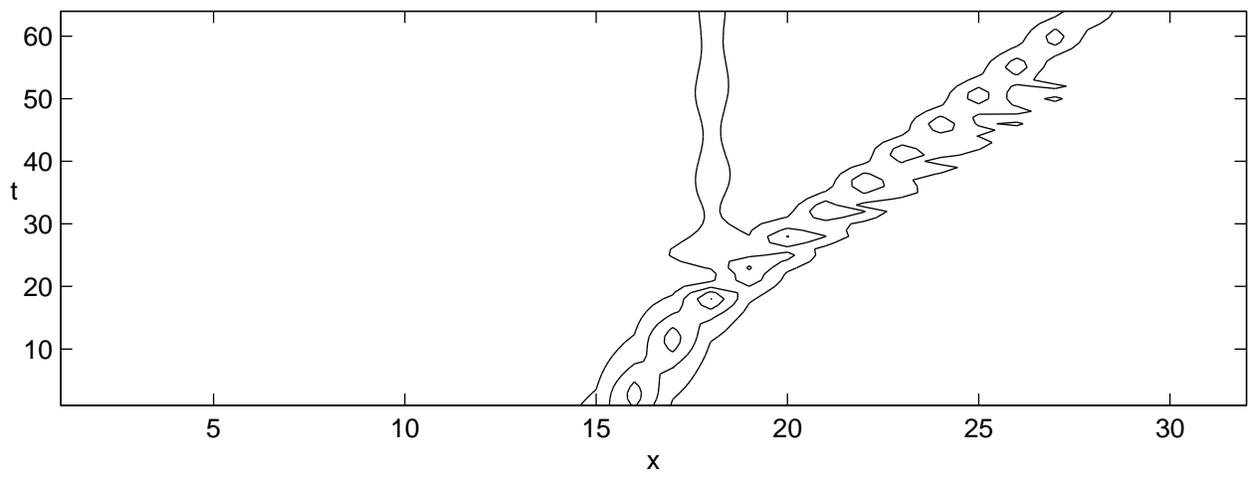
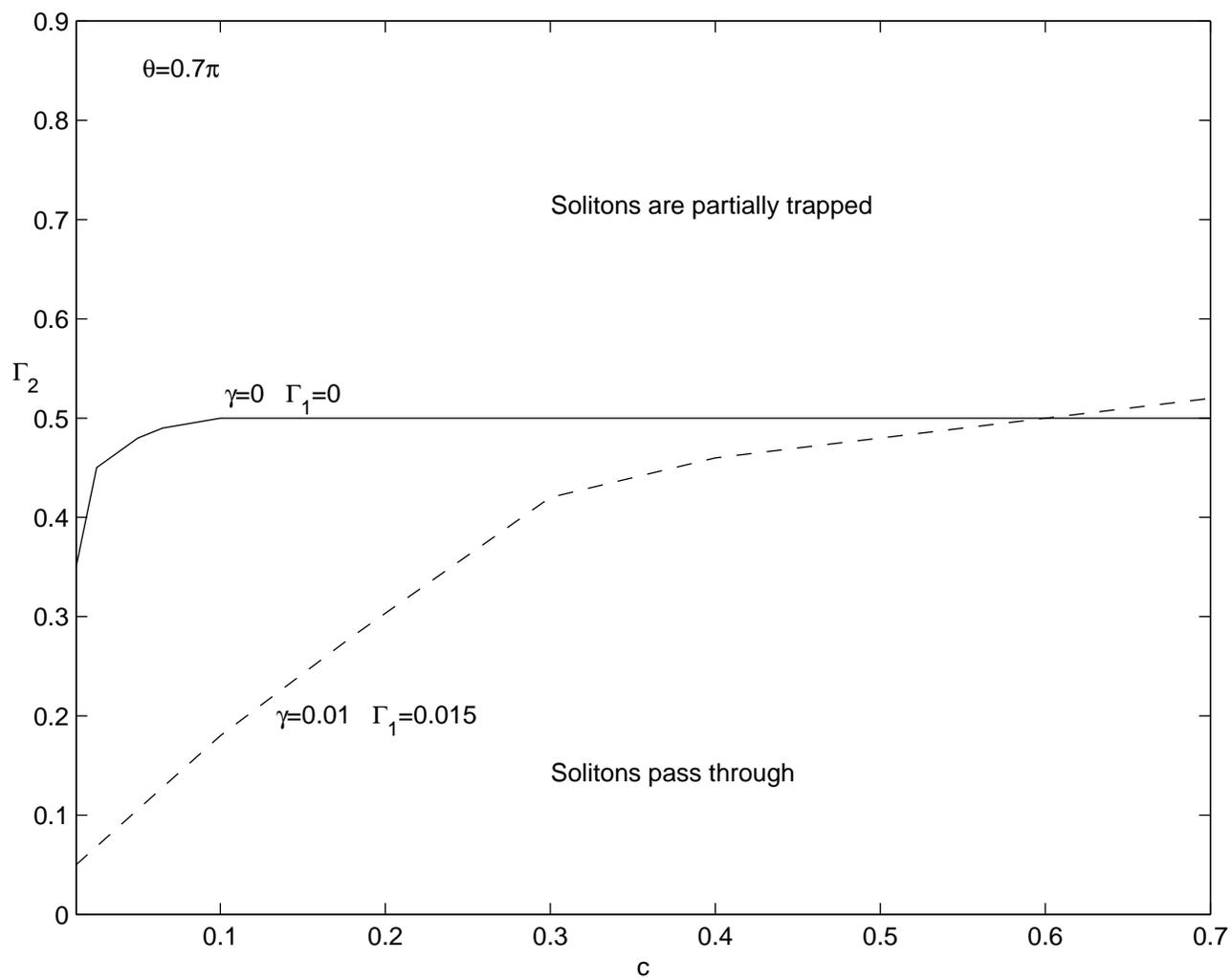
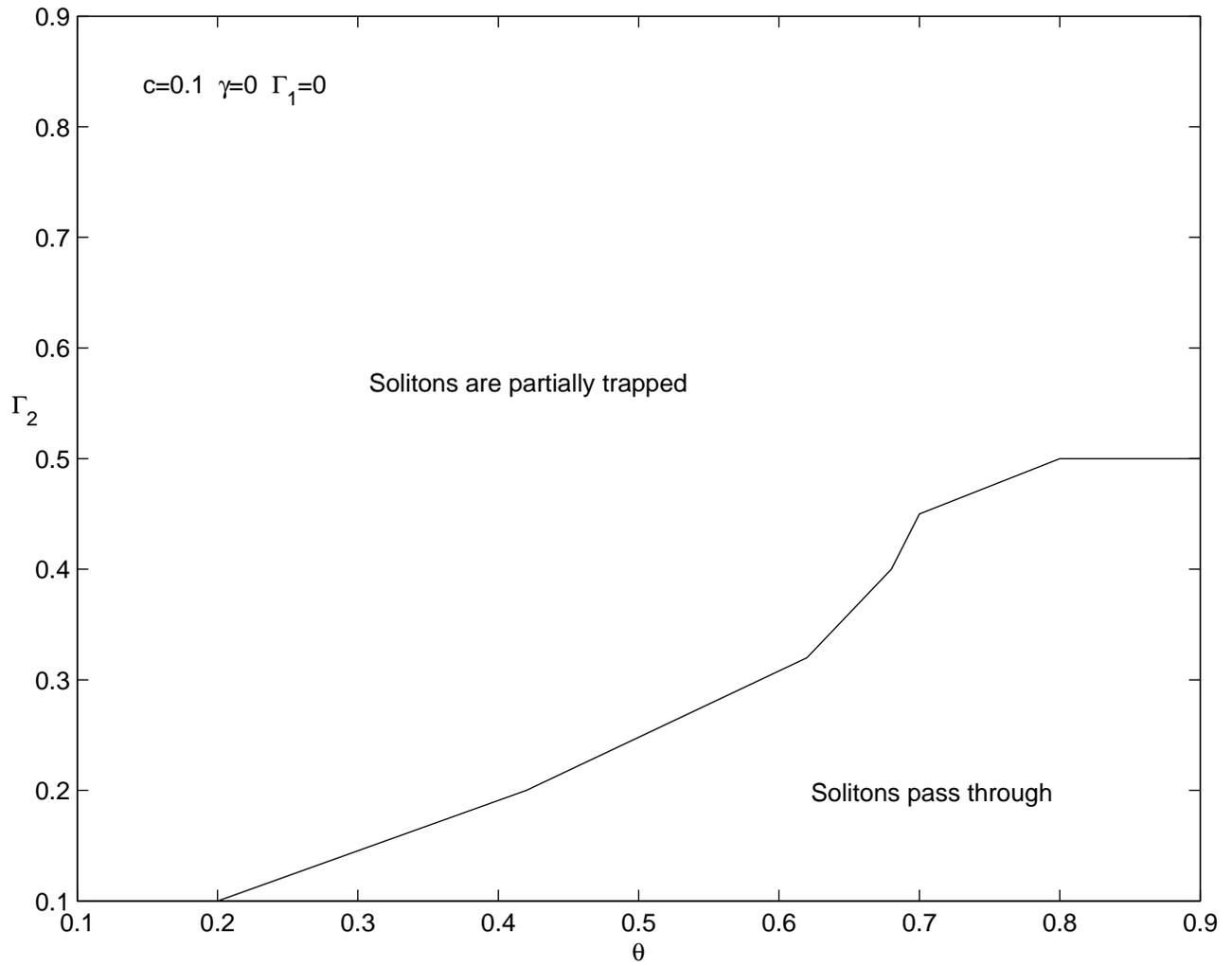
\end{document}